\documentclass[usenatbib]{ws-jai}
\usepackage[flushleft]{threeparttable}
\usepackage{amsmath,amsfonts,amssymb}
\usepackage{graphicx}
\usepackage{color}
\usepackage{array,multirow}
\usepackage[colorlinks=true,linkcolor=blue,citecolor=blue]{hyperref}

\newcommand{\paraiba}{{\it Para\'{\i}ba}}
\newcommand{\arerungua}{{\it Arerungu\'{a}}}
\newcommand{\gato}{{\it V\~ao do Gato}}
\newcommand{\urubu}{{\it Serra do Urubu}}
\newcommand{\saopaulo}{{\it S\~{a}o Paulo}}
\newcommand{\hi}{H{\sc i}}

\def\am{$^{\prime}$}
\def\as{$^{\prime\prime}$}
\begin{document}
\catchline{}{}{}{}{} 

\markboth{BINGO Collaboration}{RFI measurements and site selection}

\title{Baryon acoustic oscillations from Integrated Neutral Gas Observations:\\ Radio frequency interference measurements and telescope site selection}

\author{M.~W.~Peel$^{1*}$, C.~A.~Wuensche$^{2}$, E.~Abdalla$^{1}$, S.~Ant\'on$^{3}$, L.~Barosi$^{4}$, I.~W.~A.~Browne$^{5}$, M.~Caldas$^{6}$, C.~Dickinson$^{5}$, K.S.F.~Fornazier$^{2}$, C.~Monstein$^{7}$, C.~Strauss$^{2}$, G.~Tancredi$^{8}$, and T.~Villela$^{2}$}
\address{
${^1}$Instituto de F\'{i}sica, Universidade de \saopaulo, \saopaulo~- SP, Brazil\\
$^{2}$Divis\~{a}o de Astrof\'{\i}sica, Instituto Nacional de Pesquisas Espaciais (INPE), S\~{a}o Jose dos Campos - SP, Brazil\\
$^{3}$CIDMA, Universidade de Aveiro, Campus de Santiago, 3810-193 Aveiro, Portugal\\
$^{4}$Unidade Acad\^emica de F\'{\i}sica, Universidade Federal de Campina Grande, Campina Grande - PB, Brazil\\
$^{5}$Jodrell Bank Centre for Astrophysics, Alan Turing Building, School of Physics and Astronomy, The University of Manchester, Oxford Road, Manchester, M13 9PL, UK\\
$^{6}$MIEM, Dinatel, Sarand\'{\i} 620, 11000 Montevideo, Uruguay\\
$^{7}$Institute for Particle Physics and Astrophysics, ETH Zurich, Switzerland\\
$^{8}$Depto. Astronom\'{\i}a, Facultad de Ciencias, UdelaR, Igu\'a 4225, 11400 Montevideo, Uruguay
}
\maketitle
\corres{$^{*}$Corresponding author, email@mikepeel.net}
\begin{history}
\received{(to be inserted by publisher)};
\revised{(to be inserted by publisher)};
\accepted{(to be inserted by publisher)};
\end{history}

\begin{abstract}
The Baryon acoustic oscillations from Integrated Neutral Gas Observations (BINGO) telescope is a new \mbox{40-m~class} radio telescope to measure the large-angular-scale intensity of \hi\ emission at 980--1260\,MHz to constrain dark energy parameters. As it needs to measure faint cosmological signals at the milliKelvin level, it requires a site that has very low radio frequency interference (RFI) at frequencies around 1\,GHz. We report on measurement campaigns across Uruguay and Brazil to find a suitable site, which looked at the strength of the mobile phone signals and other radio transmissions, the location of wind turbines, and also included mapping airplane flight paths. The site chosen for the BINGO telescope is a valley at \urubu, a remote part of \paraiba\ in North-East Brazil, which has sheltering terrain. During our measurements with a portable receiver we did not detect any RFI in or near the BINGO band, given the sensitivity of the equipment. A radio quiet zone around the selected site has been requested to the Brazilian authorities ahead of the telescope construction.
\end{abstract}


\section{Introduction}
\label{sec:intro}
From a radio astronomer's viewpoint, sources of radio frequency interference (RFI) are everywhere in today's high-technology world, from personal devices, to base stations, aircraft, ship and vehicular transportation, and satellite transmissions. These signals present significant problems for radio telescopes observing at the same frequencies, particularly as radio measurements need to become increasingly more sensitive to detect ever-fainter astronomical radio signals.

The selection of a site for a new radio telescope thus significantly depends on the RFI environment---in this case the Baryon Acoustic Oscillations (BAOs) from Integrated Neutral Gas Observations (BINGO) telescope \citep{Battye2012,Battye2013,Battye2016,Dickinson2014,Wuensche2018}. BINGO will observe integrated \hi\ signals statistically in the redshift range \mbox{$z=0.13$--0.48} to constrain the properties of the Dark Sector and its possible structure \citep[]{WangAbdalla2016}. It will use the concept of ``single dish, many horns'', with a pair of 40-m mirrors and around 50 horns in a compact range layout to observe the 1420\,MHz \hi\ line emission at the redshifted frequencies of 980--1260\,MHz. This allows us to statistically observe one of the strongest spectral lines in the redshift range where dark energy becomes dominant, in a complementary way to other large-scale structure projects operating at optical or other radio frequencies.

An RFI quiet site is essential to observe the faint \hi\ signal, particularly given that this frequency range is not reserved for radio astronomy observations, instead being mainly used for aeronavigation. The optical configuration of BINGO has been chosen primarily for the clean beams with very low sidelobes to enable accurate component separation, but it will also minimize the pick-up of off-axis RFI. However, the huge sensitivity provided by the large mirrors means that RFI is still a critical issue, and one that may ultimately limit the sensitivity of the telescope.

Potential sites for a new radio telescope need to meet a number of requirements that can be quite complex, or even oppose each other. Together with accessibility and the topology of the site, RFI is a crucial part of the site selection process. RFI has been investigated for different telescope sites, particularly with strategies to identify and clean, or to flag and cut out, portions of RFI-contaminated data, including self-generated (on-site) RFI (see e.g., \citealp{2010Bentum,Offringa2013} for LOFAR; \citealp{Hidayat2014} for a possible Indonesian outrigger of SKA; \citealp{Huang2016} for the Chinese 21 Centimeter Array; \citealp{Sokolowski2016} for Murchison Radio-astronomy Observatory; and \citealp{Tao2017} for a review on RFI mitigation techniques and strategies).

In this paper we summarise the process of identifying and testing the potential sites for the BINGO telescope. Section \ref{sec:constraints} presents the constraints that the site had to meet and the potential RFI sources in the relevant frequency range. Section \ref{sec:tests} describes the equipment and method used, while Section \ref{sec:campaigns} describes the RFI measurements that were taken. Section \ref{sec:aircraft} considers interference from aircraft over the potential sites. Section \ref{sec:conclusions} concludes the paper. 

\section{Site requirements and constraints} \label{sec:constraints}
It is illustrating to calculate how bright RFI signals can be compared to the sensitivity of the BINGO telescope. An example transmitter broadcasting 100\,W (50\,dBm) of signal in a 1\,MHz channel, even located 30\,km away from the site (providing 122\,dB of free space loss at 1\,GHz), would provide around $-70$\,dBm of interference, reducing to $-130$\,dBm when seen in a $-60$\,dB sidelobe. Meanwhile the target noise level of the BINGO telescope will be around 0.1\,mK in a 1\,MHz channel, corresponding to a sensitivity of 100\,K in a $-60$\,dB sidelobe, or (assuming $\approx4$Jy/K) 400\,Jy = $4\times10^{-24}$\,W/m$^2$/Hz, which is the equivalent of $\approx-210$\,dBm, or 80\,dB fainter than the example RFI signal.

\subsection{General considerations}
The primary requirement for the site was that it has to be in an area with as low an RFI level as possible. The whole BINGO frequency range is internationally allocated to several commercial and military services. Although protected bands for radio astronomy exist, these are narrow, and are not present in the frequency range at which BINGO will operate. Our investigation has focused on sites that are at large distances from known and potential RFI sources.

The most common man-made source of interference at these frequencies is mobile phones and base-stations, which typically transmit at 700--950\,MHz. This sets the the lowest frequency for BINGO to be 980\,MHz although their harmonics can still be present in the data if not adequately filtered out. At the other end of the frequency range, radio amateurs operate at 1240--1300\,MHz (the 23\,cm band).

At 960--1215 MHz, the frequency range is reserved for aeronautical radionavigation, particularly from airplanes at \mbox{1025--1150\,MHz}. Services like secondary surveillance radar (SSR) operate at 960--1164\,MHz, and distance measurement equipment (DME) at 960--1215\,MHz. At 1215--1260\,MHz, BINGO will share the band with earth-exploration, radiolocation and radionavigation satellites, with a small share reserved for space research. In this frequency range (up to 1610\,MHz), satellites---particularly GPS-, GLONASS-, BeiDou-2- and GALILEO-navigation satellites, but also geostationary/others---will be a significant source of RFI for BINGO \citep{2003Struzak}. This potential source is site-independent and has been investigated separately by \citet{Harper2016,Harper2018}, who show that GPS satellites may provide a fundamental limitation on the sensitivity of intensity mapping experiments at these frequencies. This is because signals from GPS satellites may provide a noise floor that will not decrease as $\sqrt{t}$ --- and out-of-band emissions may contaminate all frequencies of interest, not just those where the satellite primarily transmits.

Additionally, wind turbines need to be avoided due to: 1) the high level of RFI they generate; 2) the randomness of the turbine signal orientation, which is aligned with the wind stream; and 3) the random reflection of other RFI sources by the wind turbine blades (e.g., see \citealp{Angulo2014} and references therein). With their tips extending up to 120--150\,m in height, turbine-related RFI can be detected from sources tens to hundreds of kilometres away from the telescope site, depending upon characteristics of the blades, primary emission and local topography.

Further interference can be self-generated on-site by computer equipment (wireless networking and CPU/motherboard frequencies), the telescope receivers themselves, and stray signals from calibration diodes. This class of interference is independent of site, and can be well-controlled or mitigated by a careful project design (see e.g., \citealp{Ambrosini2010}). There are also natural sources of RFI, such as lightning, weather conditions in general, which can vary between sites.

Aside from RFI, another important requirement for the site is that it should have a suitable topology for the construction of the telescope. There have been several conceptual designs for the BINGO telescope, which have changed over time (see \citealp{Battye2012,Battye2013,Dickinson2014}). The current solution considers two 40-m-class mirrors in a compact range design \citep{Dickinson2014,Wuensche2018}. Additionally, the site has to be, at the same time, reasonably accessible to researchers, technical staff, and construction teams, as well as sheltered from known and potential RFI sources using the local topography. Planning/environmental restrictions on building a telescope structure are also issues that have to be considered.

\subsection{Example RFI from Bleien Radio Observatory}
\begin{figure}
\centering
\includegraphics[width=0.75\textwidth]{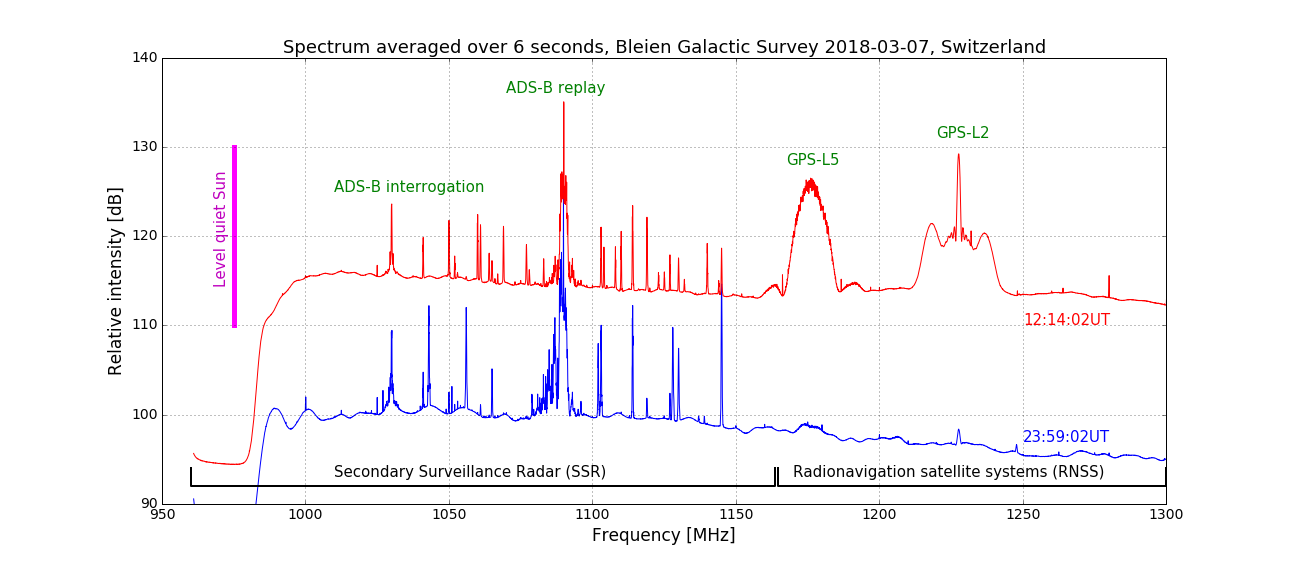}
\includegraphics[width=0.75\textwidth]{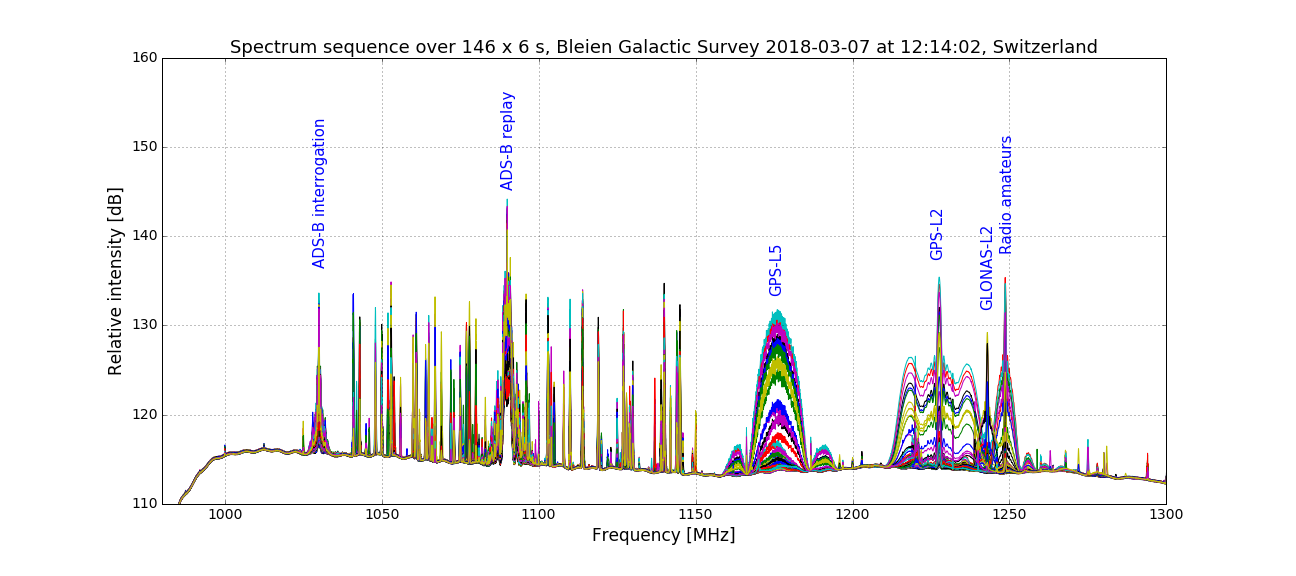}
\caption{Spectra of RFI observed with the Bleien 7\,m radio telescope. \textit{Top}: Single spectra taken over 6 seconds during the day (red line) and night (blue line). The baselines have been offset to better display the RFI signals. \textit{Bottom}: A compilation of 146 spectra each of 6 second duration. GPS-L5 and the quiet sun have similar relative intensities ($\sim$20\,dB), while GPS-L2 and ADS-B are stronger than the quiet sun at their transmission frequencies. The r.m.s. of the antenna temperature is about 0.1\,K.}
\label{fig:airplane_spectra}
\end{figure}
An example of how much RFI can be present within the BINGO frequency range is shown in Fig.\,\ref{fig:airplane_spectra}. These data, showing some measurements from Bleien Radio Observatory, Switzerland, when the RFI levels at the observatory were particularly bad, demonstrates why a remote, radio quiet area is essential for the BINGO project. The data was recorded using the Bleien 7\,m parabolic dish with a corrugated horn-antenna feeding a phase-switched pseudo-correlation receiver, followed by a heterodyne receiver and an Agilent M9703A FFT-spectrometer with 16384 frequency bins each 48.8\,kHz wide. Six filters were required to remove mobile signals, including a steep notch filter below $\sim$980\,MHz. Full details of the system are in \citet{2017Chang}. Although the site has a protection radius of 1.5\,km, this only covers nearby terrestrial sources like mobile phone, TV- or radio-transmitters, and aircraft and satellite signals are not included.

The red line in Fig.\,\ref{fig:airplane_spectra} shows a typical worst-case spectrum during daytime, with both aircraft and satellite transmissions visible in the same 6\,s measurement timespan. The blue line shows a night spectrum with less aircraft activity, and with a lower (but still visible) level of satellite emission. The baselines have been offset to better display the RFI signals. The magenta vertical line at the left side shows a typical intensity step when the telescope points to the quiet sun.

\section{RFI testing}\label{sec:tests}

\subsection{Equipment}
For the measurements on the candidate BINGO sites, we used an omni-directional discone antenna with a low-noise amplifier connected directly to the antenna, followed by a long low-loss cable to a portable spectrum analyzer. This provides reasonable sensitivity to RFI from any azimuthal direction, and is a standard setup used for these tests \citep[see e.g.,][]{Fonseca2006,ITU2011}. The method is similar to the SKA protocol for site RFI characterization \citep{Ambrosini2003}, but is slightly less rigorous as it was necessary to keep the time per measurement shorter.

In most Brazilian campaigns, a commercial logarithmic periodic dipole array (LPDA) antenna was also used in the measurements to explore RFI from specific directions, although we do not show any measurements from it in this work. For operational reasons, the exact components used had to be replaced through the testing stages. On-site measurements were performed in Uruguay and Brazil with two different spectrum analyzers: an Anritsu S331E spectrum analyser was used in Uruguay, while an Agilent N9343C spectrum analyzer was used in Brazil (and \arerungua, Uruguay).

\begin{wstable}[t]
\centering
	\caption{Details of the omnidirectional discones used for the RFI measurements.}
	\begin{tabular}{@{}ccc@{}}
	\toprule
		{\bf Model} & D130J & Custom \\
        \colrule
		{\bf Bandwidth} & 25-1300\,MHz & 250--2600\,MHz \\
		{\bf Center frequency} & 660\,MHz & 1450\,MHz\\
		{\bf Gain} & 2\,dBi & \\
		{\bf Disc diameter} & 560\,mm & 102.8\,mm  \\
		{\bf Cone narrow end diameter} & & 1.84\,mm  \\
		{\bf Cone broad end diameter} & & 146.8\,mm \\
		{\bf Cone side length} & 820\,mm & 146.8\,mm  \\
		{\bf Antenna mouth diameter} & & 146.8\,mm \\
		{\bf Antenna height} & & 128.2\,mm  \\
		{\bf Cone angle} & & 29.59$^\circ$  \\
        \colrule
        {\bf Notes} & Uruguay & Brazil\\
	\botrule
	\end{tabular}
	\label{tab:antennas}
\end{wstable}

\begin{figure}
   \centering
   \includegraphics[width=0.25\linewidth]{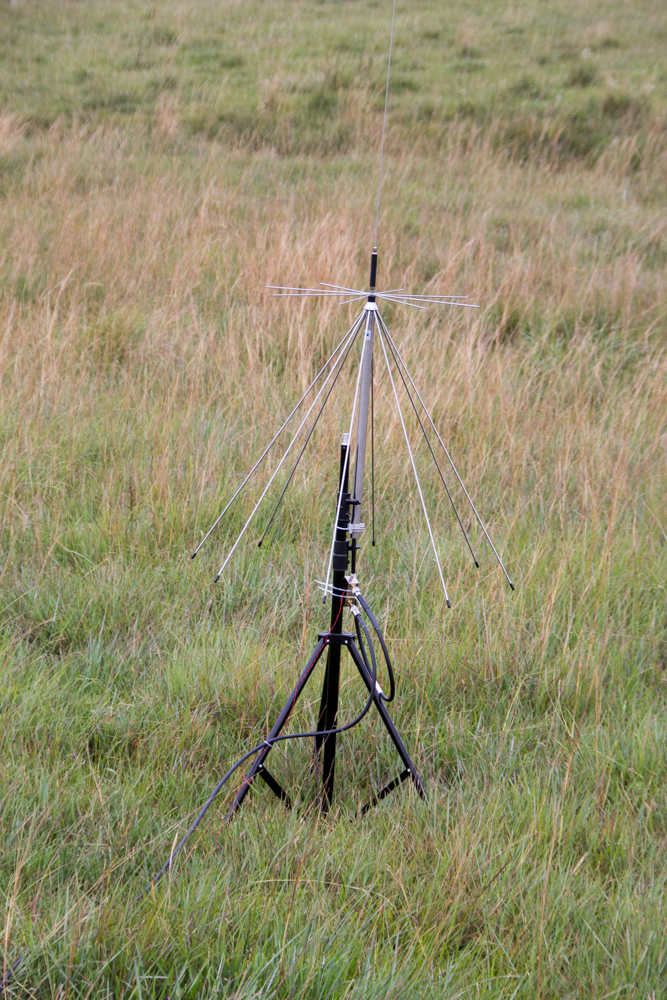}
   \includegraphics[width=0.25\linewidth]{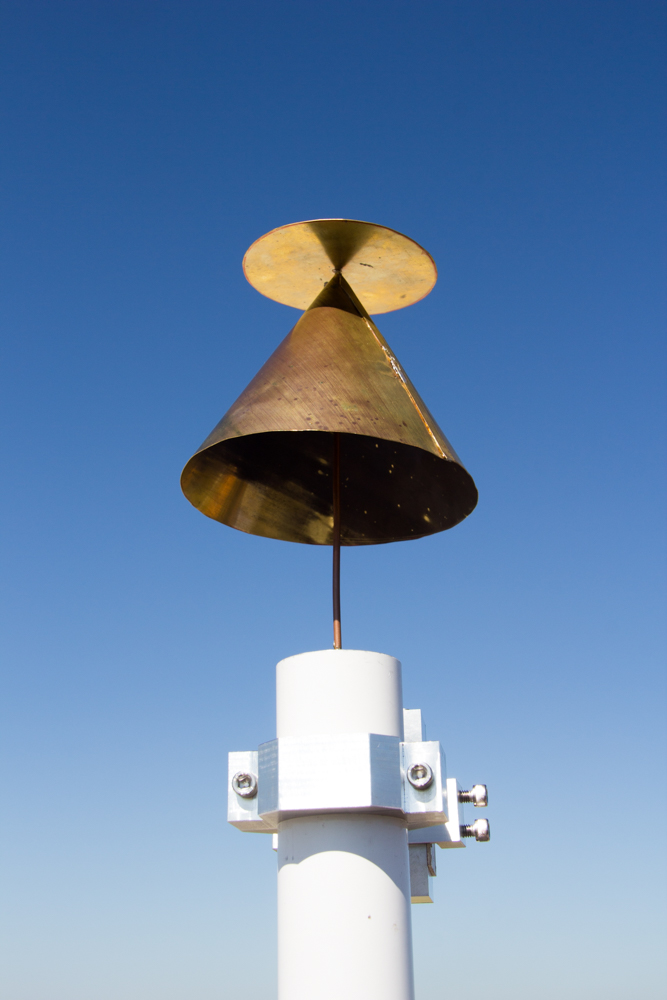}
   \caption{The discone antennas used in this work. Left: D130J wire discone. Right: custom solid discone.}
   \label{fig:discone_pictures}
\end{figure}
\begin{figure}
   \centering
   \includegraphics[width=0.48\linewidth]{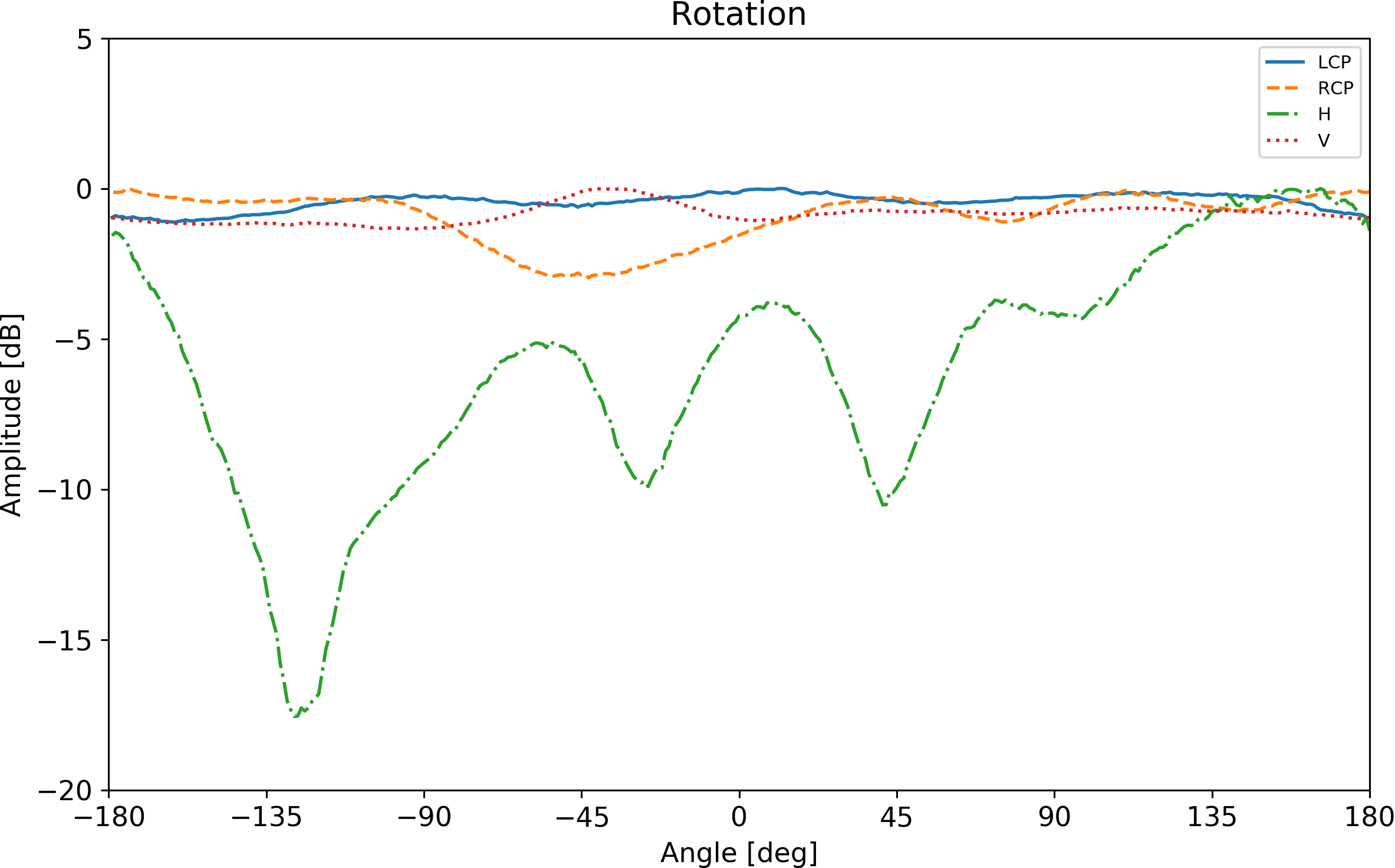}
   \includegraphics[width=0.48\linewidth]{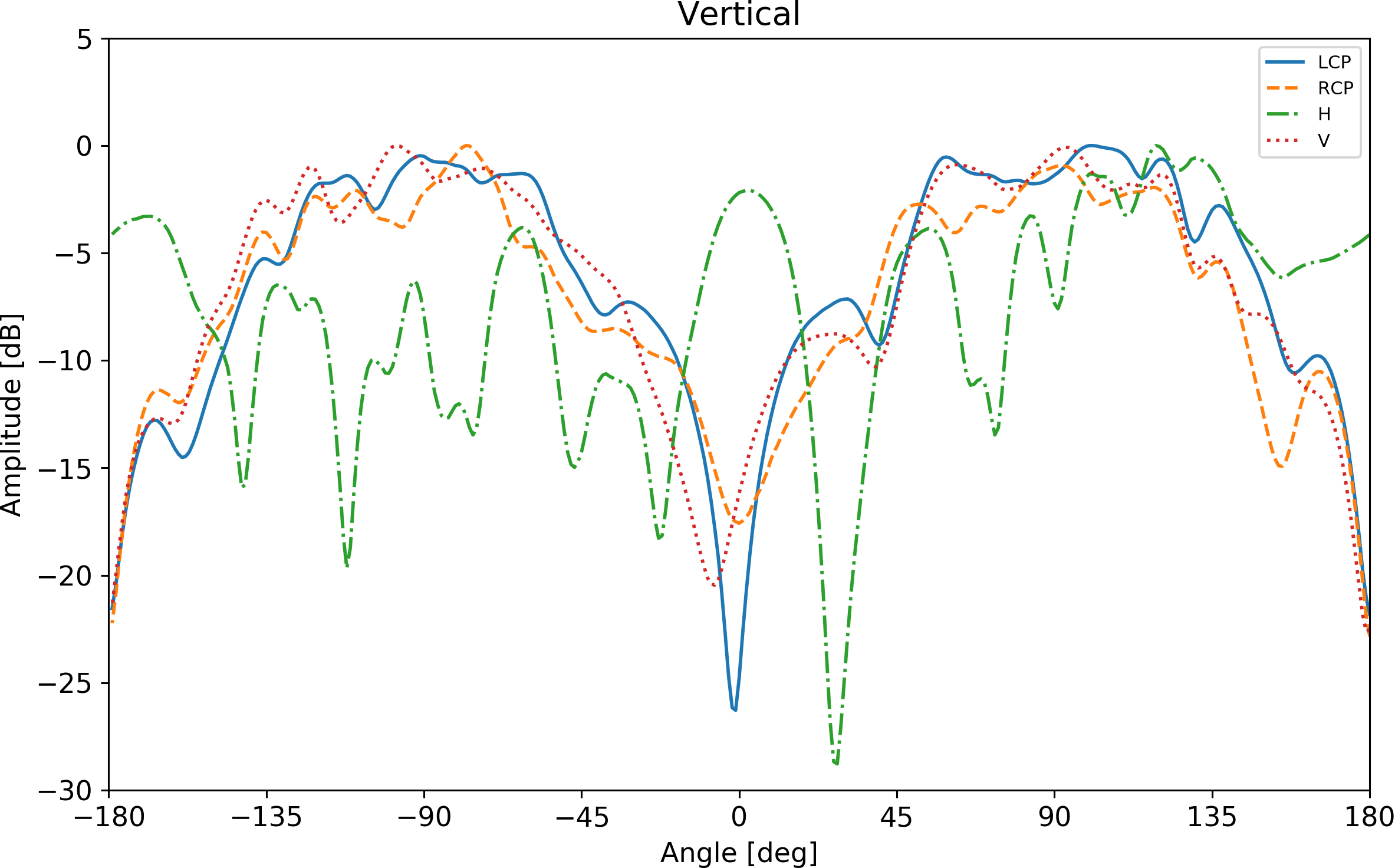}
   \caption{The measured performance of the custom discone antenna at 1\,GHz when rotated (left) and angled from the vertical (middle) with left-circular (blue solid) right-circular (orange dashed), horizontal (green dot-dashed) and vertical (red dotted, polarization. All lines have been normalized so that their maximum value is at 0\,dB. The discone has an irregular response to horizontally-polarised emission, but a good response to vertical and circular polarisation.}
   \label{fig:discone}
\end{figure}

Three antennas were used in this campaign: the first discone was used in Uruguay, the second discone and the LPDA in Brazil. The characteristics of the discones can be found in Table\,\ref{tab:antennas}, and they are shown in Figure~\,\ref{fig:discone_pictures}. One of the discone antennas and the LPDA are commercially-available and have publicly-available standard responses\footnote{\url{https://www.diamondantenna.net/d130j.html}, \url{http://proeletronic.com.br/produto/antena-celular-quadband/}}. The other discone antenna was custom-built at INPE to cover the BINGO frequency range along with lower and higher frequencies, and had its performance measured at the horn testing facility at INPE's Laboratory of Integration and Tests. The results when the antenna is rotated and angled from the vertical are shown in Fig.\,\ref{fig:discone}. INPE's discone has a consistent response to vertical and circularly-polarised emission, and a worse response to horizontal polarisation and to emission above and below the discone, as expected given the geometry. This means that it is only sensitive up to $45^{\circ}$ elevation  from horizontal, so it will not detect signals from airplanes and satellites passing overhead.

\begin{wstable}[t]
\caption{Amplifiers used for the RFI measurements} 
\begin{tabular}{@{}cccccc@{}}
\toprule
 & Miteq & Kuhne & Aaronia & Miteq & Minicircuits  \\
\colrule
{\bf Model} & AFS3-00100300-12-10P-4 & KU 1090A (custom) & AG UBBVX & AFS3 & ZX60-P33ULN+ \\
{\bf Frequencies} & 100--3000\,MHz & 960--1260\,MHz & Unknown & 100--3000\,MHz & 400-3000\,MHz \\
{\bf Gain} & 34\,dB & 30\,dB & Unknown & 30\,dB & 15\,dB \\
{\bf Noise Figure} & 1.2\,dB & 0.5\,dB & Likely 3+\,dB & 1.8\,dB & 0.4\,dB \\
\colrule
{\bf Notes}  & Uruguay \& Brazil & Uruguay \& Brazil & \paraiba & \paraiba & pair used in series \\
 & (initial) & (initial) & (initial) & (middle) & (final measurements)\\
\botrule
\end{tabular}
\label{tab:amplifiers}
\end{wstable}

The amplifiers were used in the order presented in Table \ref{tab:amplifiers}. We used the LNAs on their own, and also with the pre-amplifier in the spectrum analyser on, although all results shown in this paper did not use the pre-amplifier as the first amplifier provided sufficient gain and the pre-amplifier introduced some internally-generated spectral features into the measured spectra. Measurements were made either using the `max-hold' function on the spectrum analyzer, in order to capture transient signals, or (for the Agilent N9343C spectrum analyser) by recording a waterfall of data and then post-processing this to calculate both the maximum and mean signals recorded. Calibration measurements were taken by replacing the antenna with a $50~\Omega$ load with a known (measured) physical temperature $T_\mathrm{cal}$.

\subsection{Method} 
\label{sec:method}
In order to obtain calibrated spectra we followed the procedures by \citet{Kuzmin1966} for receiver respective antenna calibration. These procedures are based on the knowledge of the receiver temperature $T_\mathrm{rx}$ or of the receiver noise figure (NF). In our case $T_\mathrm{rx}$ was of the order of 100\,K, given a dedicated low noise amplifier with a noise figure of less than 1\,dB (including coaxial cable, connectors, antenna loss etc.). Two measurements are required, the first one of which is a measurement of a $50~\Omega$ termination resistor at known ambient temperature $T_\mathrm{cal}$, leading to a power spectrum denoted to $P_\mathrm{ref}[\mathrm{dBm}]$. Usually, $T_\mathrm{cal}$ is assumed to be of the order of 290\,K. If the ambient temperature is measured, then $T_\mathrm{cal}$ can be put into the equation~\ref{eq1} to give a more precise result. Taking into account noise contribution from the calibration source, we can estimate the noise figure (NF) of the spectrum analyzer by
\begin{equation}
\mathrm{NF}[\mathrm{dB}] = P_\mathrm{ref}[\mathrm{dBm}] - 10\log_{10}\left(\frac{k~T_\mathrm{cal}~\mathrm{RBW}}{0.001\mathrm{W}}\right) \label{eq1}.
\end{equation}
Here, $k$ is the Boltzmann constant and RBW denotes the spectrum analyzer radiometric bandwidth, which in our case was set to either 1\,MHz or 0.3\,MHz.

A second observation of the radio frequency interference is performed with the discone antenna, leading to a power spectrum denoted by $P_\mathrm{obs}[\mathrm{dBm}]$ that contains both the signal generated by RFI and the receiver noise of the spectrum analyzer. Based on these two measurements we can then evaluate the RFI power $P_\mathrm{RFI}[\mathrm{dBm}]$:
\begin{equation}
P_\mathrm{RFI}[\mathrm{dBm}] = P_\mathrm{obs}[\mathrm{dBm}] - \mathrm{NF}[\mathrm{dB}] \label{eq2}.
\end{equation}
Ideally the radiometer equation would also be taken into account by subtracting $\sqrt{N\tau\,\mathrm{VBW}} [\mathrm{dB}]$, where N is the number of measurements, $\tau$ the integration time for each measurement, and VBW is the video bandwidth of the spectrum analyser. However, this is only correct in the absence of RFI signals, so it has been omitted here.

Flux density information is computed using the effective area of the receiving antenna $A_\mathrm{eff}$. According to ITU regulation ITU-R RA.769-1 \citep{ITU2003} the antenna gain G is defined to be 0\,dB ($G=1$), because we never know from which direction and at which polarization the RFI waves are coming from. So, the antenna effective area is calculated according to:
\begin{equation}
A_\mathrm{eff} = G~\frac{\lambda^2}{4\pi} =\frac{\lambda^2}{4\pi}.
\label{eq4}
\end{equation}

Combining the previous equations we obtain the radio flux density \citep{Kraus1965} of the incoming RFI according to:
\begin{equation}
S[\mathrm{W/m^2/Hz}] = \frac{8~\pi~P_\mathrm{RFI}[\mathrm{W}]}{\lambda^{2}~\mathrm{RBW}}.
\label{eq5}
\end{equation}
We calculate both the mean and the maximum signals received at each frequency; the latter uses the maximum of the calibration signal to account for the noise bias that would otherwise be present.

The flux density $S$ derived in equation \ref{eq5} helps to study the effects of RFI-like cross-modulation, instrument saturation and data loss as defined in \citet{ITU2015}, but has no effect in terms of protection as none of the BINGO band is protected.

\section{Measurement campaigns}\label{sec:campaigns}
\begin{figure}
   \centering
   \includegraphics[width=0.6\linewidth]{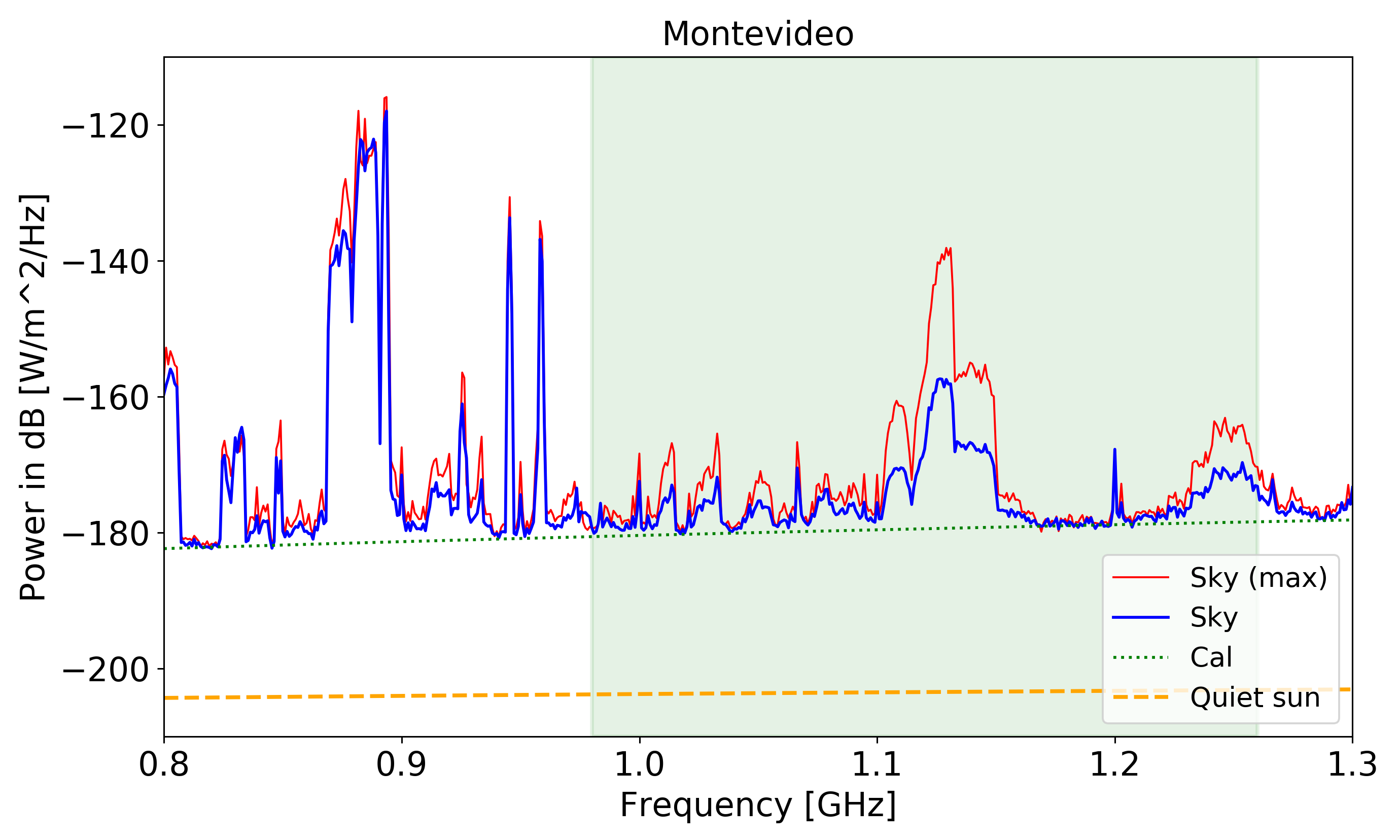}
   \caption{RFI measurement at University de La Republica, Montevideo, using a discone antenna. The maximum signal recorded is shown in red; the mean is shown in blue; the dark green dotted line shows the calibration signal; and the orange dashed line shows the level of the quiet sun. The BINGO frequency band of 980--1260\,MHz is highlighted in green. Mobile phone signals are present at 800--950\,MHz, as well as in-band signals from unknown sources. Note that the y-axis in this plot extends to $-100$\,dB due to the brightness of mobile phone signals in the city, while in later plots it is cut off above $-140$\,dB to show the fainter signals more clearly.}
   \label{fig:montevideo}
\end{figure}

A number of RFI measurement campaigns were undertaken in Uruguay and Brazil to survey potential BINGO sites. An example of the type of spectra obtained during these measurements is shown in Figure\,\ref{fig:montevideo}; this is for an urban setting at University de La Republica, \textit{Montevideo}, Uruguay, to clearly demonstrate RFI in the spectrum. In this figure, the blue line shows the average signal received by the spectrum analyser, with the red line showing the maximum signal received at each frequency. The green dotted line shows the level of the calibration signal. RFI can be seen at most frequencies, with mobile phone signals particularly visible at 800--950\,MHz, as well as signals within the BINGO band (green shaded region), particularly at 1.12\,GHz (from an unknown source) and at 1.2\,GHz (possibly a radio link). For comparison, the level of emission from the quiet sun is shown in yellow, based on a statistical model by \citet{Benz1999}. Later plots in this section follow the same style.

\begin{wstable}[t]
\caption{Selected sites where the RFI campaigns were undertaken.}
\begin{tabular}{@{}cccc@{}}
\toprule
\# & Name & Coordinates & Notes \\
\colrule
U1 & {\it Castrillon}, Uruguay & 31$^\circ$31\am19\as S, 55$^\circ$29\am40\as W & Land ownership problems\\
U2 & \arerungua, Uruguay & 31$^\circ$39\am58\as S, 56$^\circ$34\am13\as W & Isolated slope in military field\\
B1 & {\it S\~ao Martinho}, Brazil &  29$^{\circ}$26\am24\as S, 53$^{\circ}$48\am39\as W & INPE observatory. Mobiles, radar.\\
B2 & {\it Cachoeira Paulista}, Brazil & 22$^{\circ}$21\am41\as S, 44$^{\circ}$59\am02\as W & INPE observatory. Mobiles, aircraft.\\
B3 & {\it Parque dos Dinosauros} &  06$^\circ$44\am04\as S, 38$^\circ$15\am47\as W & Scientific reserve\\
B4 & {\it Cruzeiro de Pianc\'o} & 07$^\circ$01\am57\as S 37$^\circ$52\am10\as W & Isolated area\\
B5 & \gato, Brazil & 07$^\circ$03\am00\as S, 38$^\circ$37\am57\as W & Isolated area, in the middle of ridges, \\
   &               &                                                &  support from federal university\\
B6 & \urubu, Brazil & 07$^\circ$02\am57\as S, 38$^\circ$15\am46\as W & Isolated area, in the middle of ridges, \\
   &               &                                                 & support from federal university\\
\botrule
\end{tabular}
\label{tab:sites}
\end{wstable}

\subsection{Uruguay campaign}
\begin{figure}
    \begin{center}
        \includegraphics[width=0.48\linewidth]{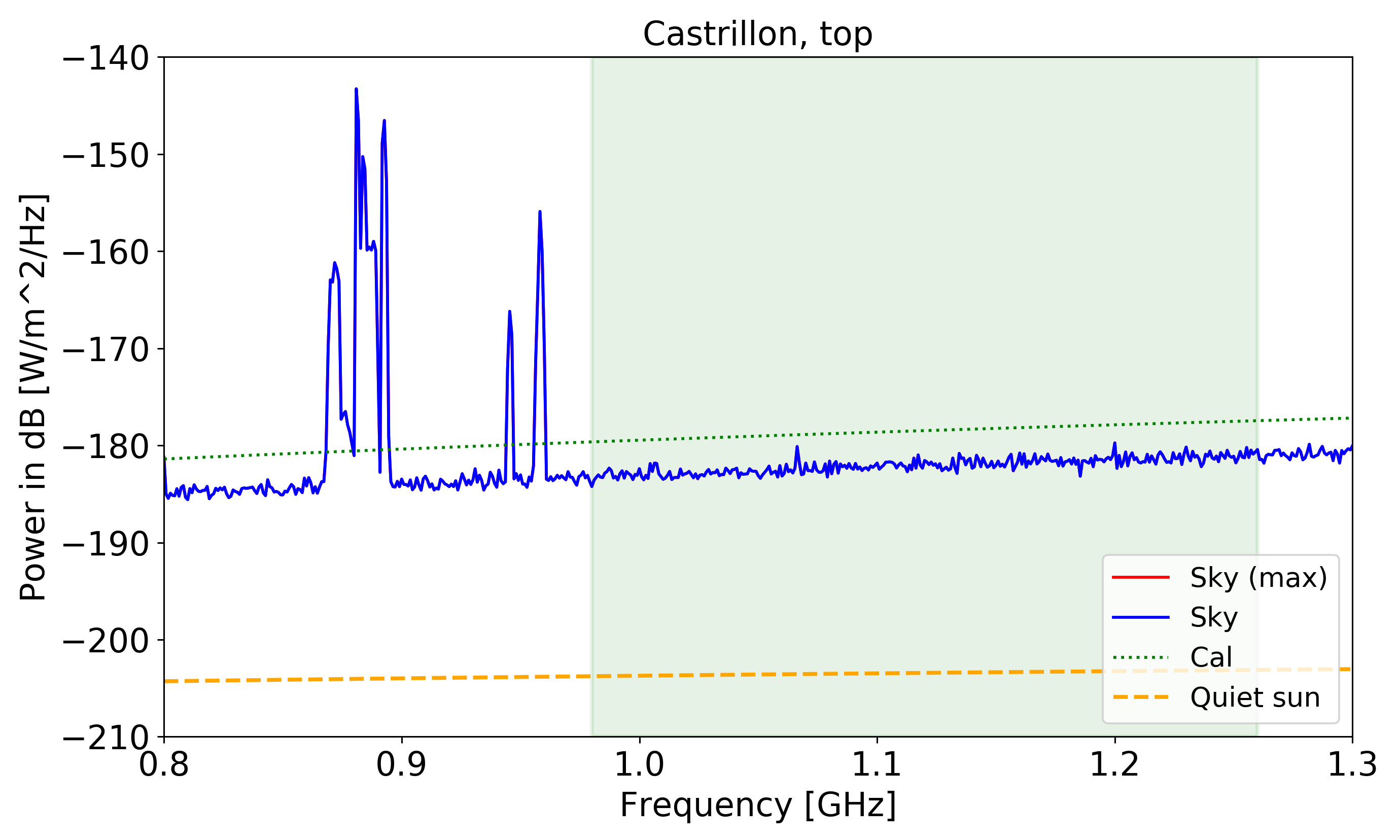}
        \includegraphics[width=0.48\linewidth]{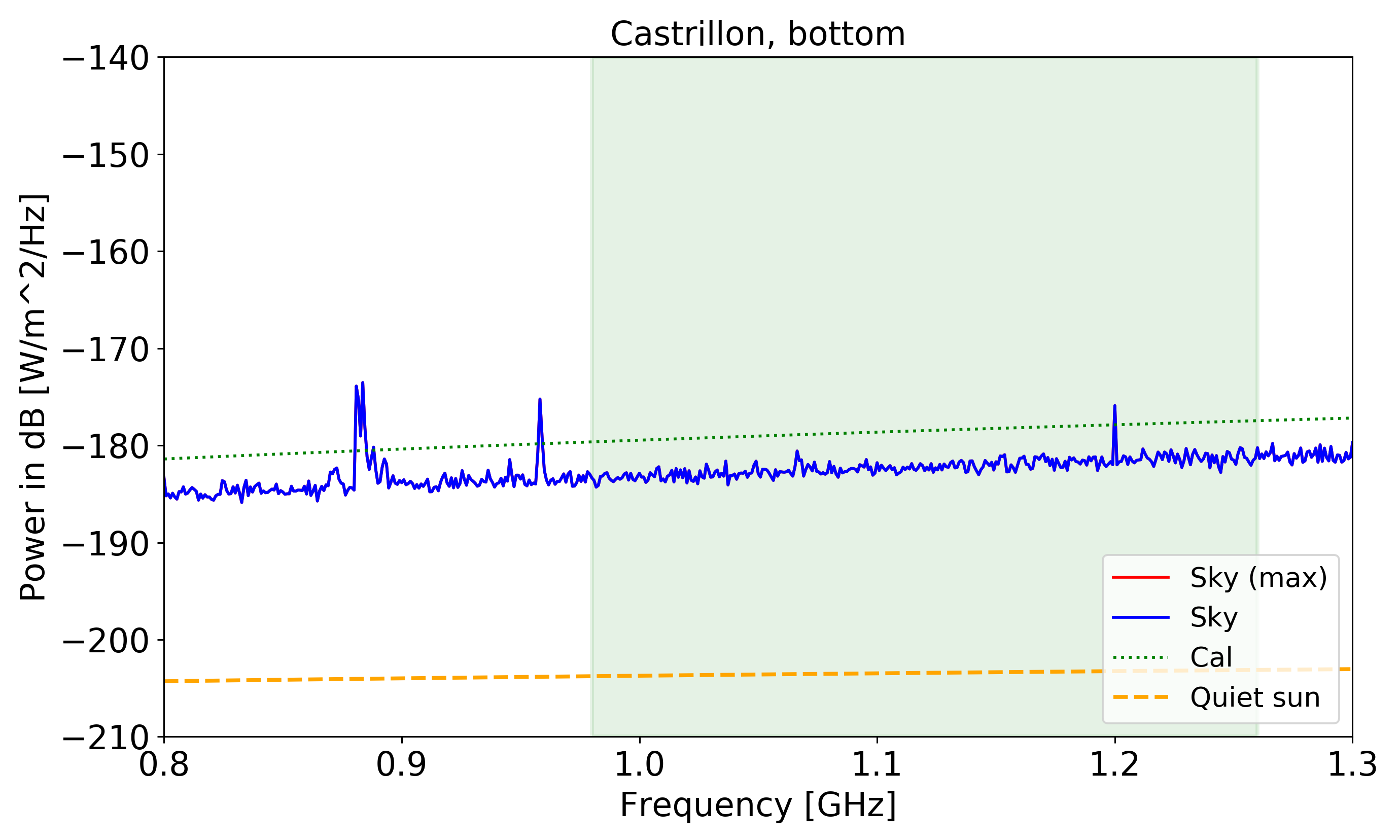}
   \includegraphics[width=0.48\linewidth]{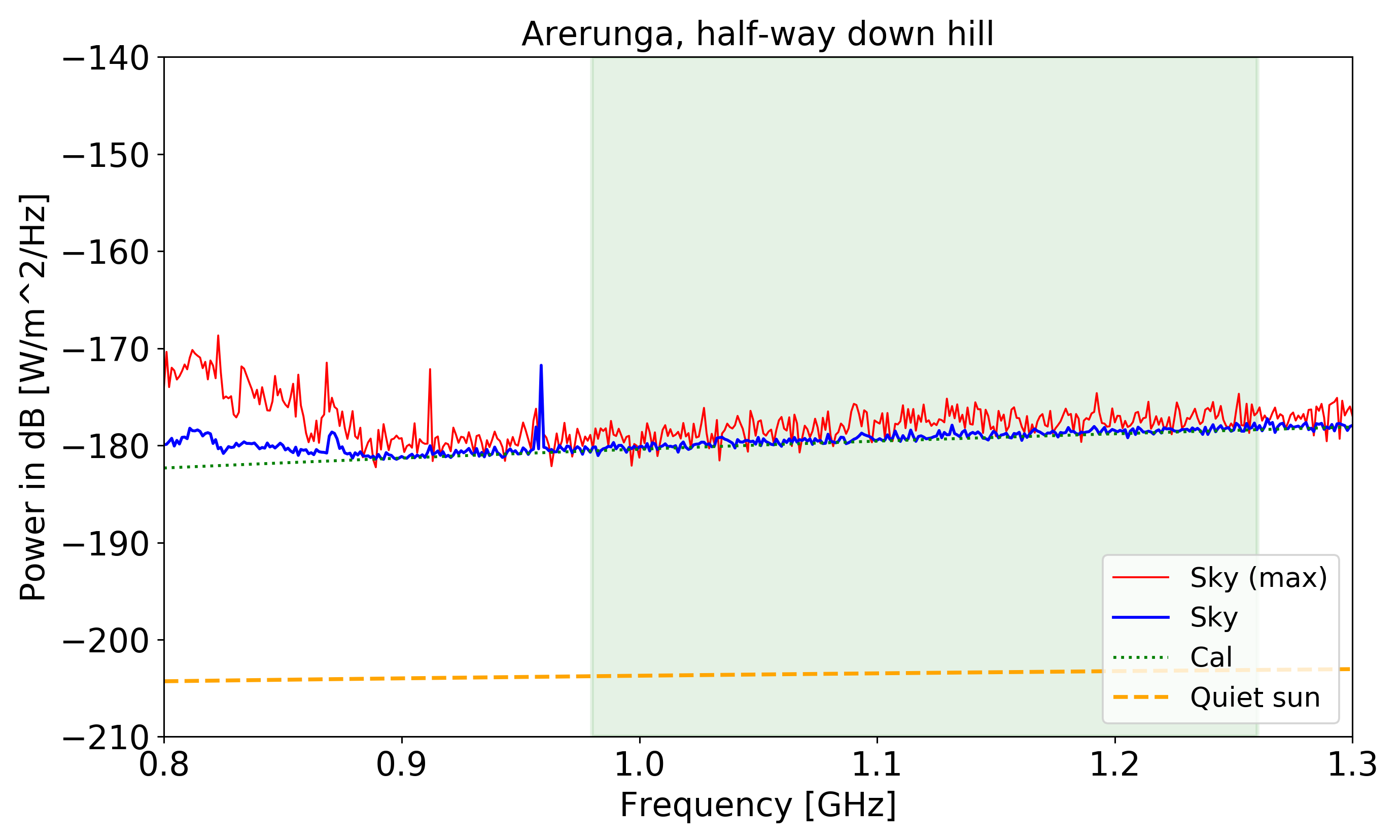}
   \includegraphics[width=0.48\linewidth]{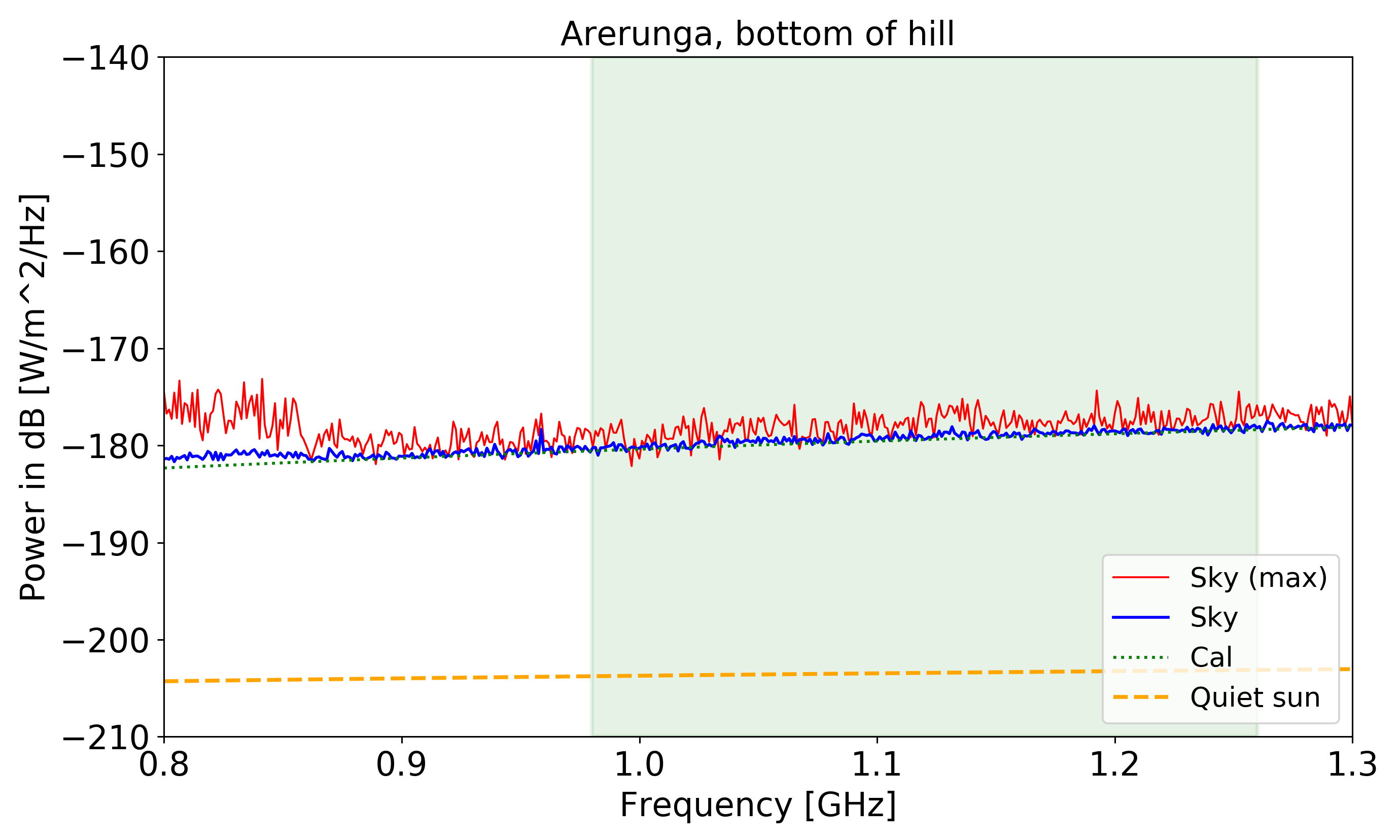}
        \caption{RFI calibrated measurements at the top ({\it top-left}) and bottom ({\it top-right}) of the Castrillon mine, and at the middle ({\it bottom-left}) and bottom ({\it bottom-right}) of the slope in \arerungua. Colours are as per Fig.~\ref{fig:montevideo}. Only maximum values are available for the Castrillon measurements. Note the cell phone peaks at around 900 and 950\,MHz, and the signal at 1.2\,GHz at Castrillon.}
        \label{fig:uruguay}    
    \end{center}
\end{figure}
The first BINGO RFI campaign was carried out in Uruguay in November 2013, focusing on remote areas in the Rivera Province with suitable topology for the telescope structure. Eight locations were surveyed, including a number of quarries (El Portugues, La Bulldog, Knob Hill and Castrillon), valleys (Arroyo Pelotas and Paso del Cerro), as well as existing observatories (Aiguas and Los Molinos). From these, an abandoned gold mine in Castrillon, near the town of Minas de Corrales, was selected (``U1'' in Table \ref{tab:sites}), as the topography was well suited to the support of the telescope structure and the RFI environment above the mobile phone frequencies was reasonable \cite{Dickinson2014,Battye2016}. However, mobile phone signals were visible at 850--950\,MHz, and later measurements revealed additional signals at higher frequencies such as 1.2\,GHz (see Fig.~\ref{fig:uruguay}). This option was ruled out in late 2016 due to negotiation difficulties with the land owner.

An alternative site was subsequently identified at \arerungua\ (``U2'' in Table \ref{tab:sites}), where a military field used for raising cattle was available. Two RFI campaigns were conducted in \arerungua\ in November 2016 and March 2017. The RFI environment of the remote site was acceptable, despite having a few high power windmills less than 100 km away. A suitable slope was identified at this site, and measurements of middle and bottom of the slope are shown in Fig. \ref{fig:uruguay}. The RFI environment at the bottom of the slope was the best that we had seen so far, and this became the baseline site for the telescope.

The initial BINGO designs used reference horns to look at an Celestial Pole and provide a constant reference signal for a correlation receiver \citep{Battye2013}. This design put an effective requirement to locate the telescope at a latitude further than $30^{\circ}$ from the equator. Some sites were investigated in the northern hemisphere, particularly in the UK, before considering South America as possibility, with the countryside of Uruguay as target. However, a number of technical and financial constraints removed this requirement. Due to this, as well as constraints on the ways that existing funding could be spent, alternative sites closer to the equator in Brazil were investigated.

\subsection{Brazil campaign}
\begin{figure}
    \begin{center}
        \includegraphics[width=0.48\linewidth]{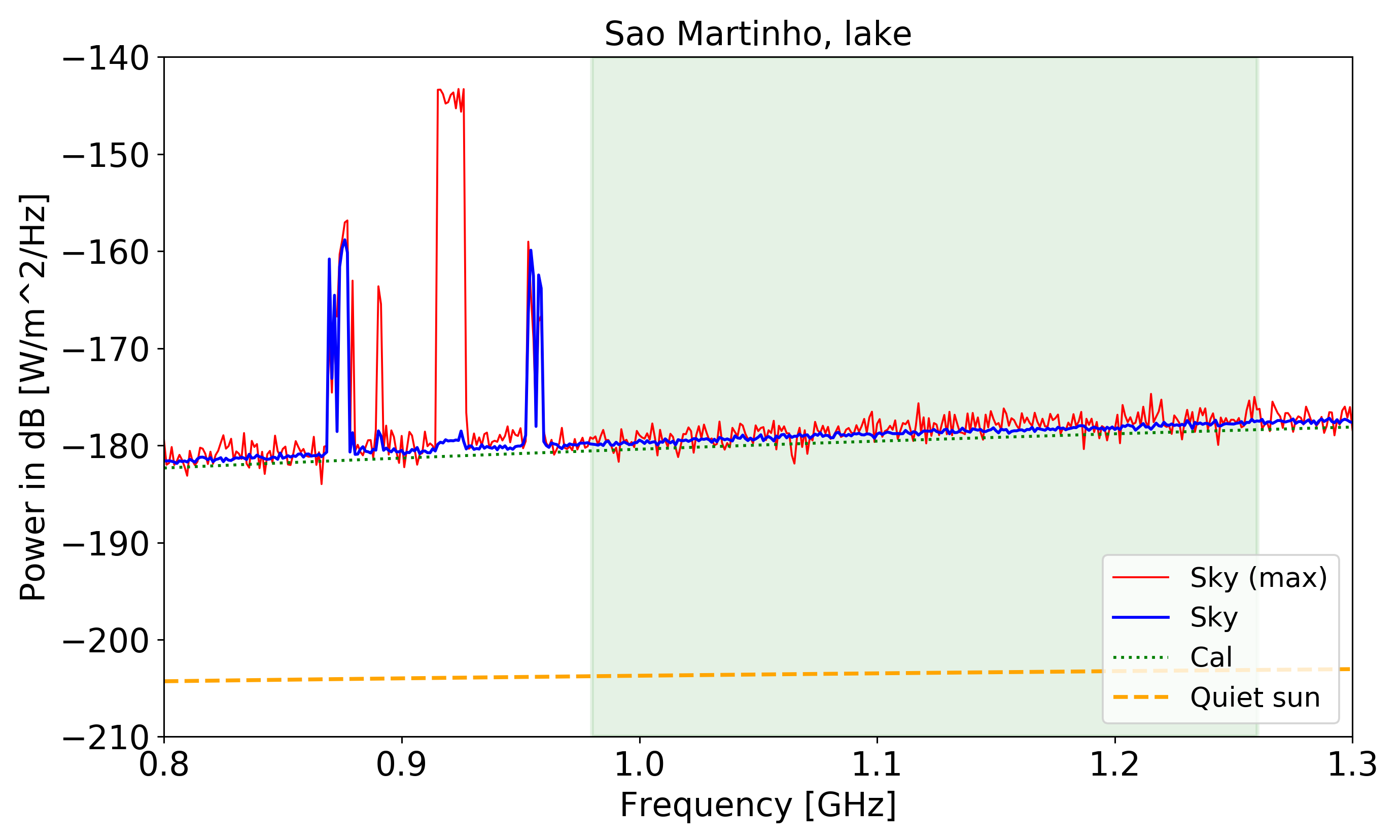}
        \includegraphics[width=0.48\linewidth]{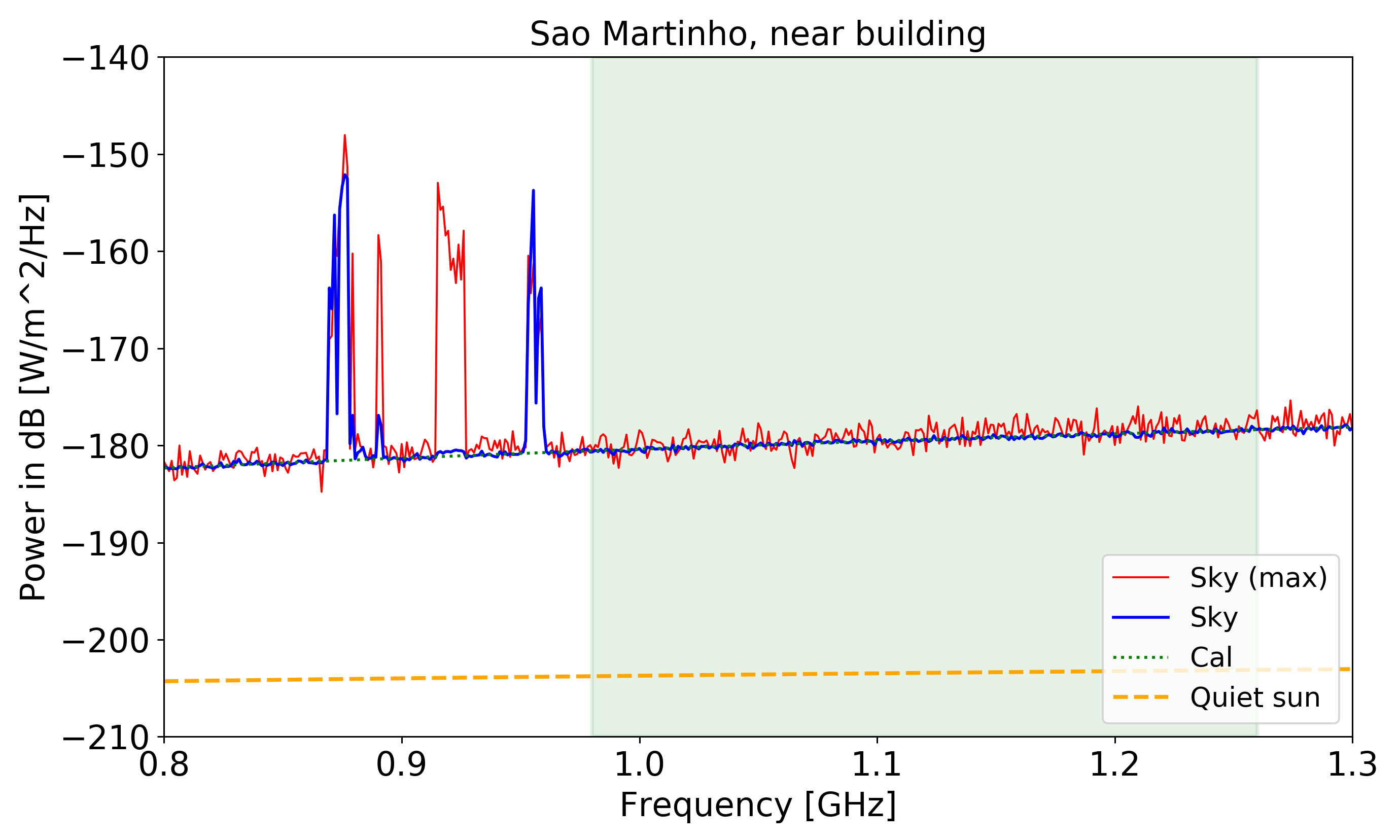}
        \includegraphics[width=0.48\linewidth]{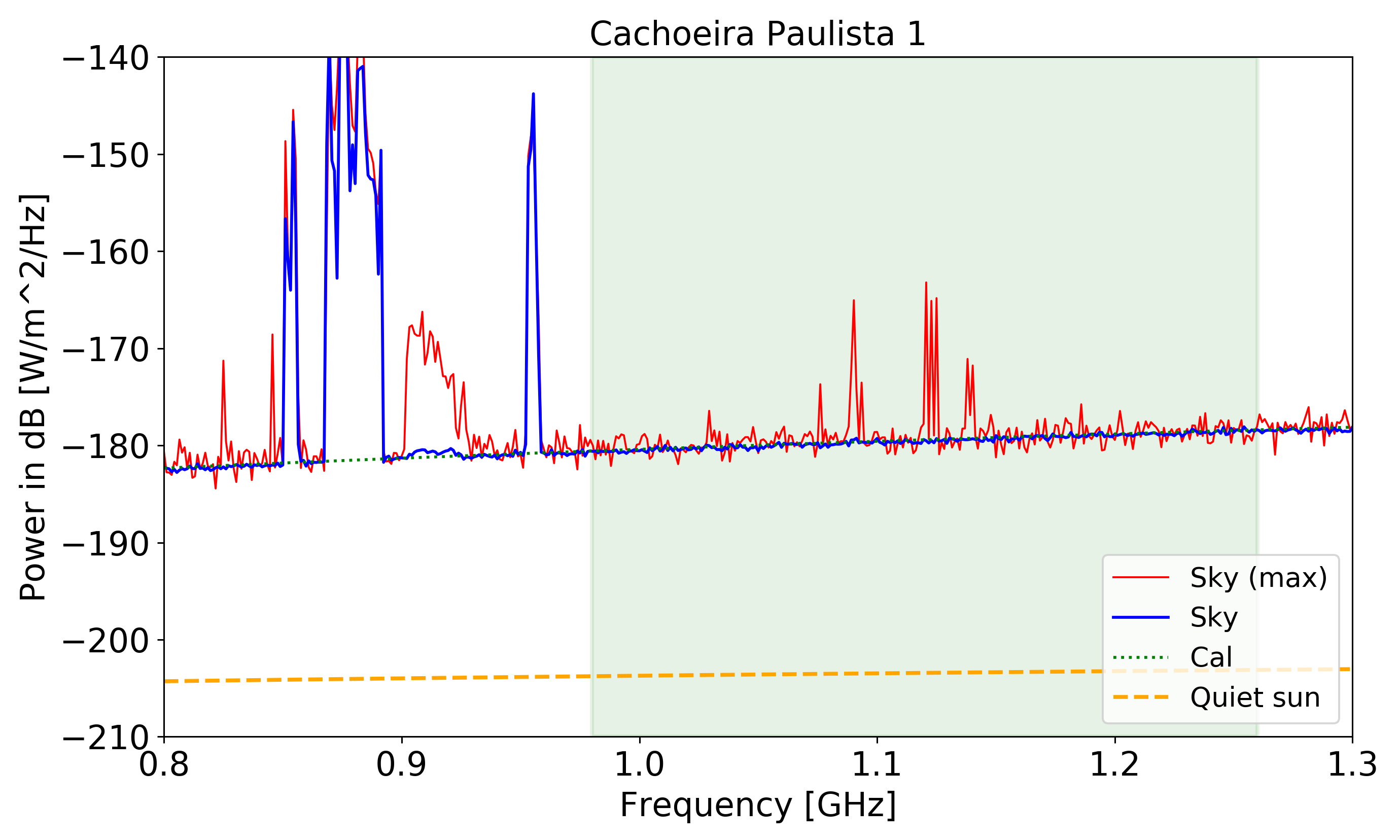}
        \includegraphics[width=0.48\linewidth]{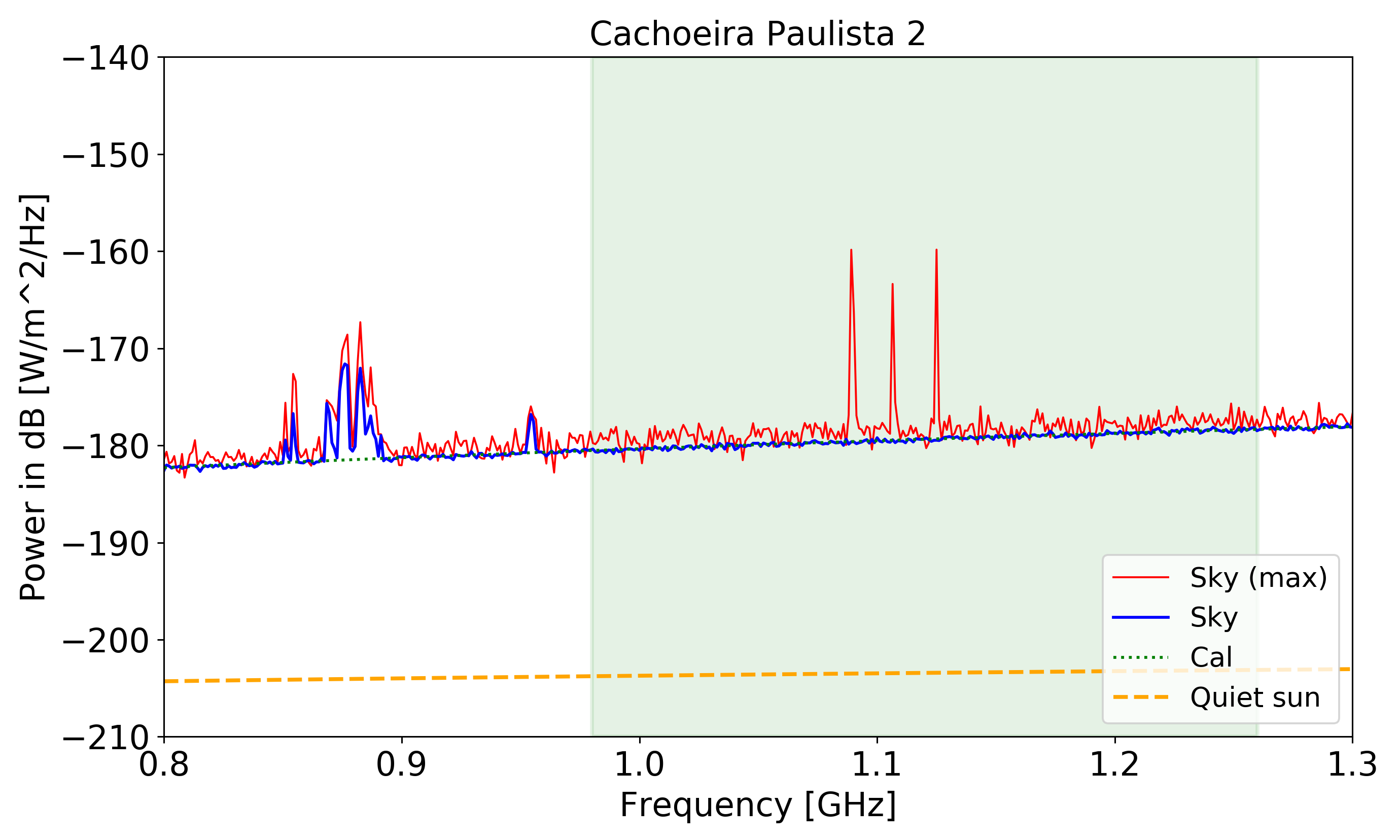}
    	\caption{Calibrated 10-minute RFI measurements using the discone antenna. {\it Top}: At INPE Observat\'{o}rio Espacial do Sul (RS) near the observatory lake (left) and on top of a lab building (right). Strong mobile phone signals are seen at 860--960\,MHz. {\it Bottom:} at INPE Cachoeira Paulista. Measurements in two different spots, in hidden areas surrounded by cliffs of about 60\,m. Note again the mobile phone lines in the same band, as well as in-band signals from aircraft. Colours are as per Fig.~\ref{fig:montevideo}.}
	 \label{RFI_SM}
    \end{center}
\end{figure}

Initially, sites in the south of Brazil were looked at. The first candidate was the Observat\'{o}rio Espacial do Sul (``B1'' in Table \ref{tab:sites}), an INPE facility near \textit{S\~ao Martinho da Serra} in \textit{Rio Grande do Sul} state. Two campaigns were conducted in 5--8 February 2017 and 20--21 March 2017 using first the custom discone antenna with the first Miteq amplifier, and then the commercial discone with the Kuhne amplifier for direct comparison to Arerungua. Two measurements from these campaigns are shown in Fig.~\ref{RFI_SM}, where strong mobile phone signals can be seen at 970 and 950\,MHz, along with a strong transient signal at 920\,MHz. Despite having very good infrastructure, these results, along with the presence of airport and military radars around $\sim$ 40\,km away, disqualified the site.

The next option (``B2'' in Table \ref{tab:sites}) was another INPE facility in \textit{Cachoeira Paulista}, \saopaulo\ state, where three campaigns were conducted during April and May 2017 using both the custom discone and the LPDA, with the Kuhne amplifier. The RFI level there was also high, as can be seen in Fig. \ref{RFI_SM}, with strong mobile phone signals, in-band interference, and many radio and TV transmitters in the town. Additionally, the airspace between the cities of \saopaulo\ and \textit{Rio de Janeiro} is very busy. The same overall conclusions reached in \textit{S. Martinho da Serra} also applied in this case: despite having useful infrastructure, the very high RFI level rules out this site.

\begin{figure}
	\centering
	\includegraphics[width=0.95\linewidth]{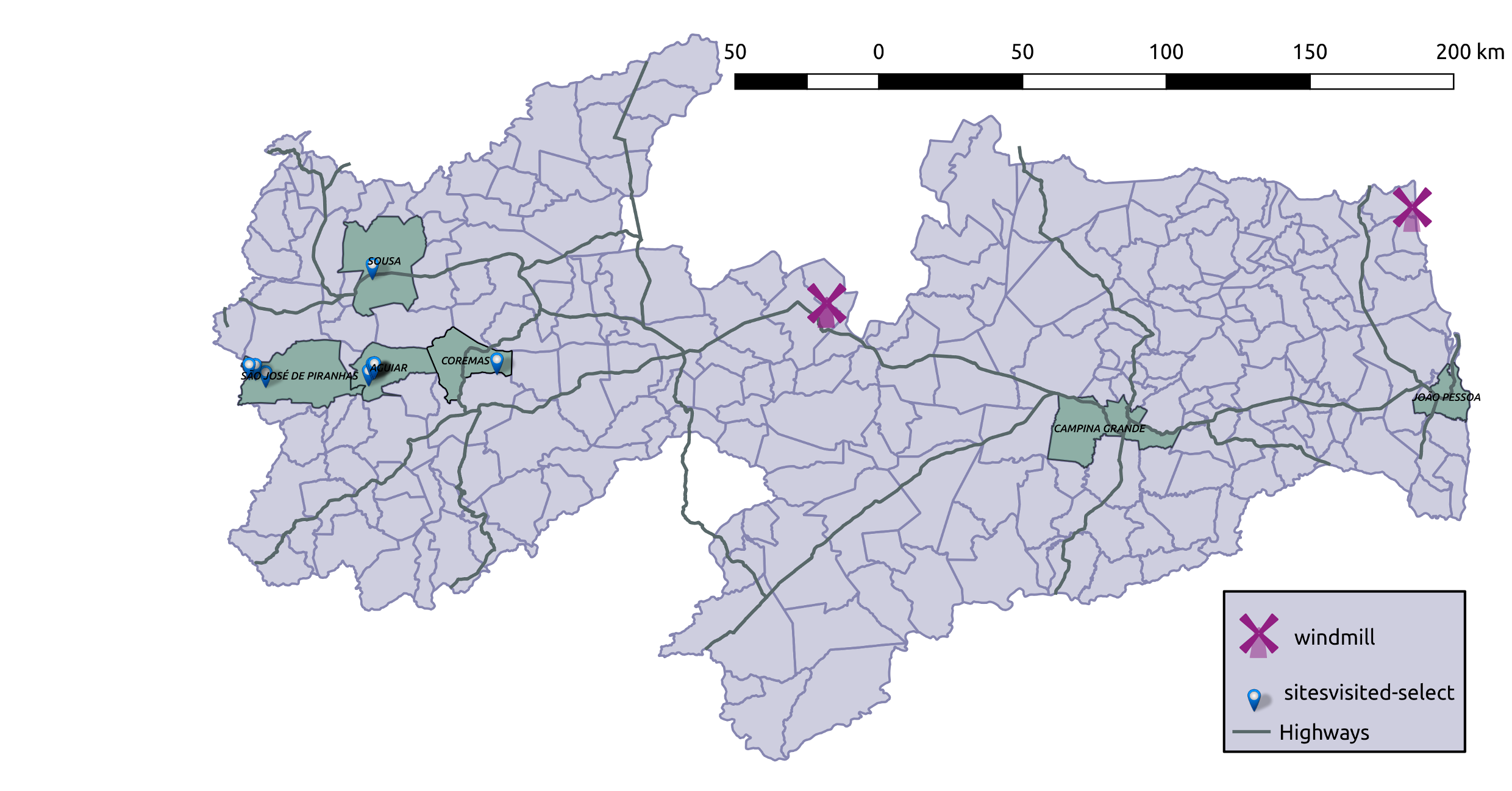}
	\caption{A map of \paraiba\ indicating the most relevant measurement sites. Relevant cities are highlighted and windmills depicted.}
	\label{fig:ERBs3}
\end{figure}

More isolated regions further north were then investigated in May and early June 2017. Previous studies were performed in \textit{Goi\'as} for a Square Kilometre Array RFI campaign and \textit{Bahia} for a preliminary site investigation (without RFI measurements) to install GNSS stations for geological and deep space network tracking purposes. A contact with the Universidade Federal de Campina Grande opened a new window of possibilities to search for sites with a low level of RFI in the state of \paraiba, in the north-east of Brazil. Eleven sites were visited in three different campaigns that took place in June, July and August 2017, and their locations are shown in the map in Fig. \ref{fig:ERBs3}.

\begin{figure}
\centering
\includegraphics[width=0.5\linewidth]{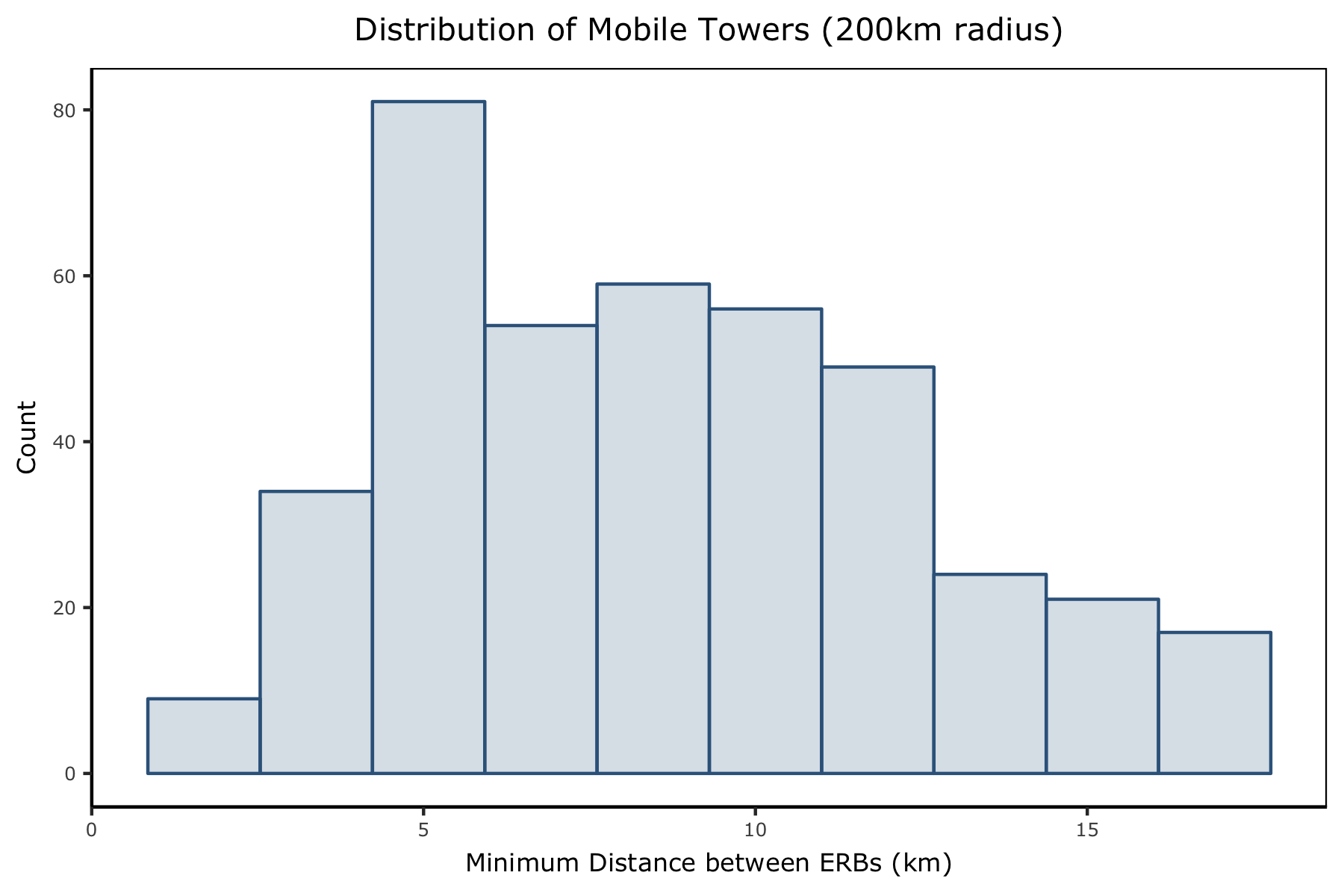}
\caption{Distribution of minimum distances between mobile towers.}
\label{fig:ERBs1}
\end{figure}

If we consider the record of mobile phone towers installed in a radius of 200\,km around Aguiar, western Paraiba,\footnote{available from \url{http://www.telebrasil.org.br/panorama-do-setor/mapa-de-erbs-antenas}} we can see that the towers are densely spaced. Their average separation is 8.7\,km (see the distribution in Fig.~\ref{fig:ERBs1}), so we can only expect to find a clear RFI zone in sheltered valleys.

For all measurements, we used the directional and the custom discone antennas described above. Initial measurements used the Kuhne amplifier, before issues with that led us to use an Aaronia amplifier temporarily, then a Miteq amplifier with 1.8\,dB noise figure. The final measurements were made using the Minicircuits LNA, which produced the most sensitive measurements of the three campaigns.

\begin{figure}
   \centering
   \includegraphics[width=0.48\linewidth]{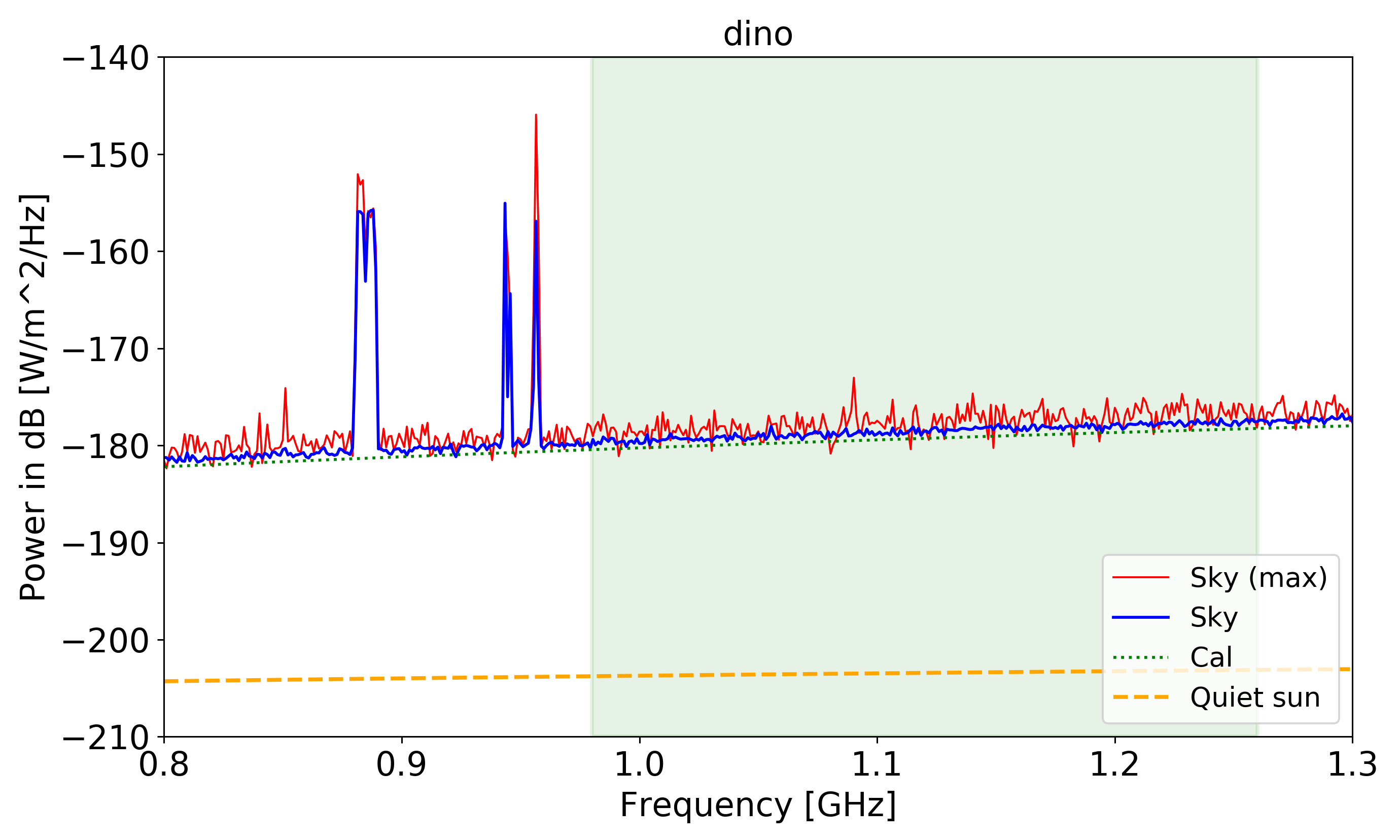}
   \includegraphics[width=0.48\linewidth]{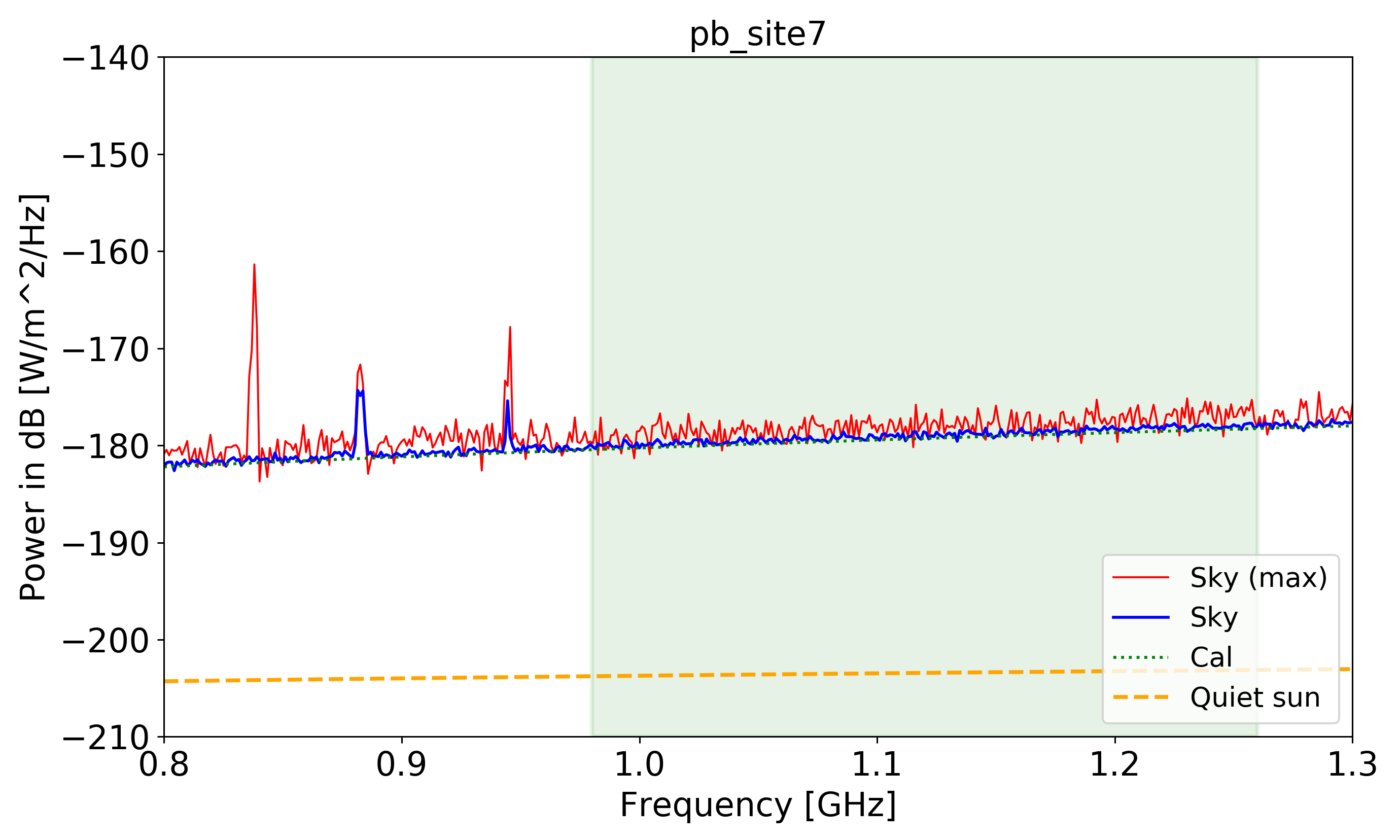}
   \includegraphics[width=0.48\linewidth]{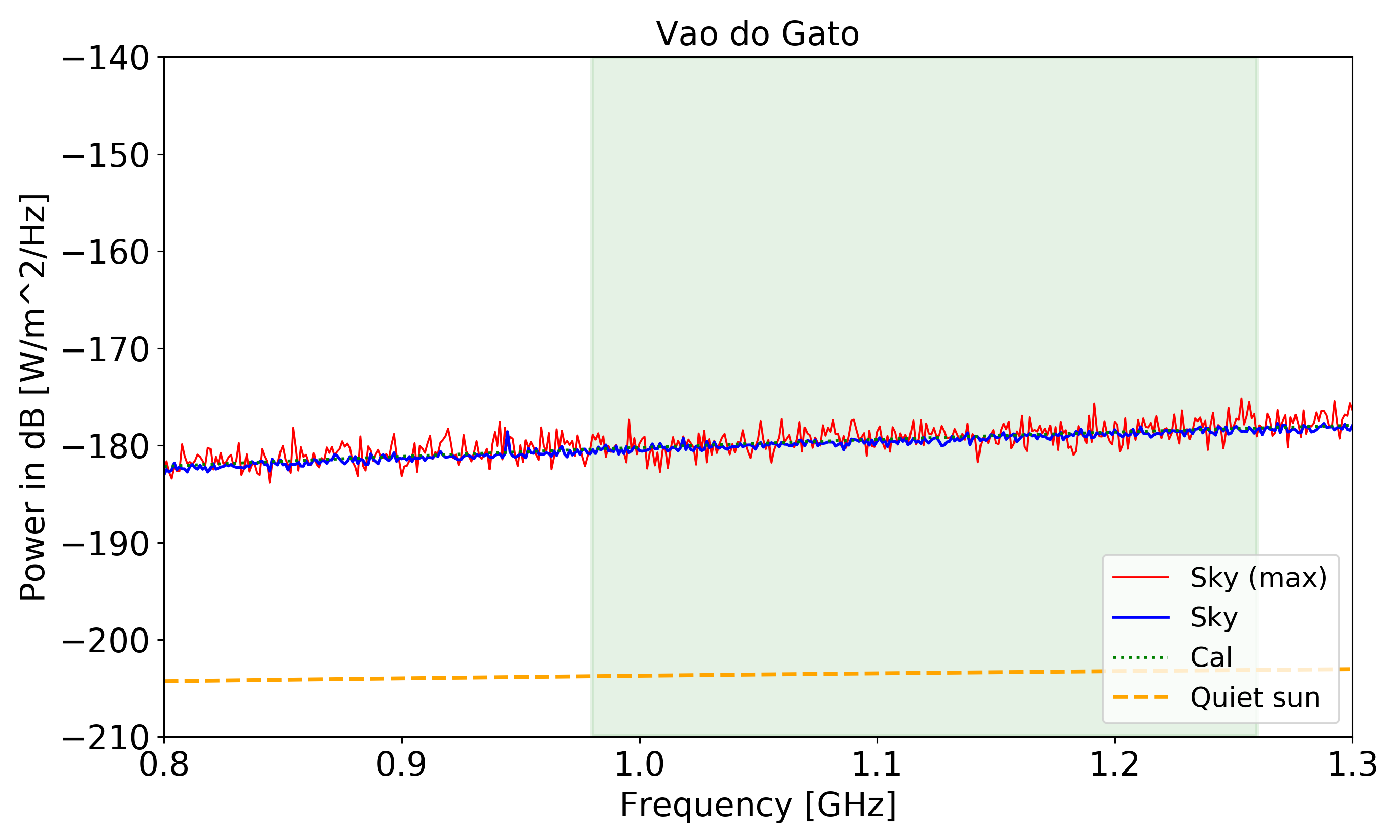}
   \includegraphics[width=0.48\linewidth]{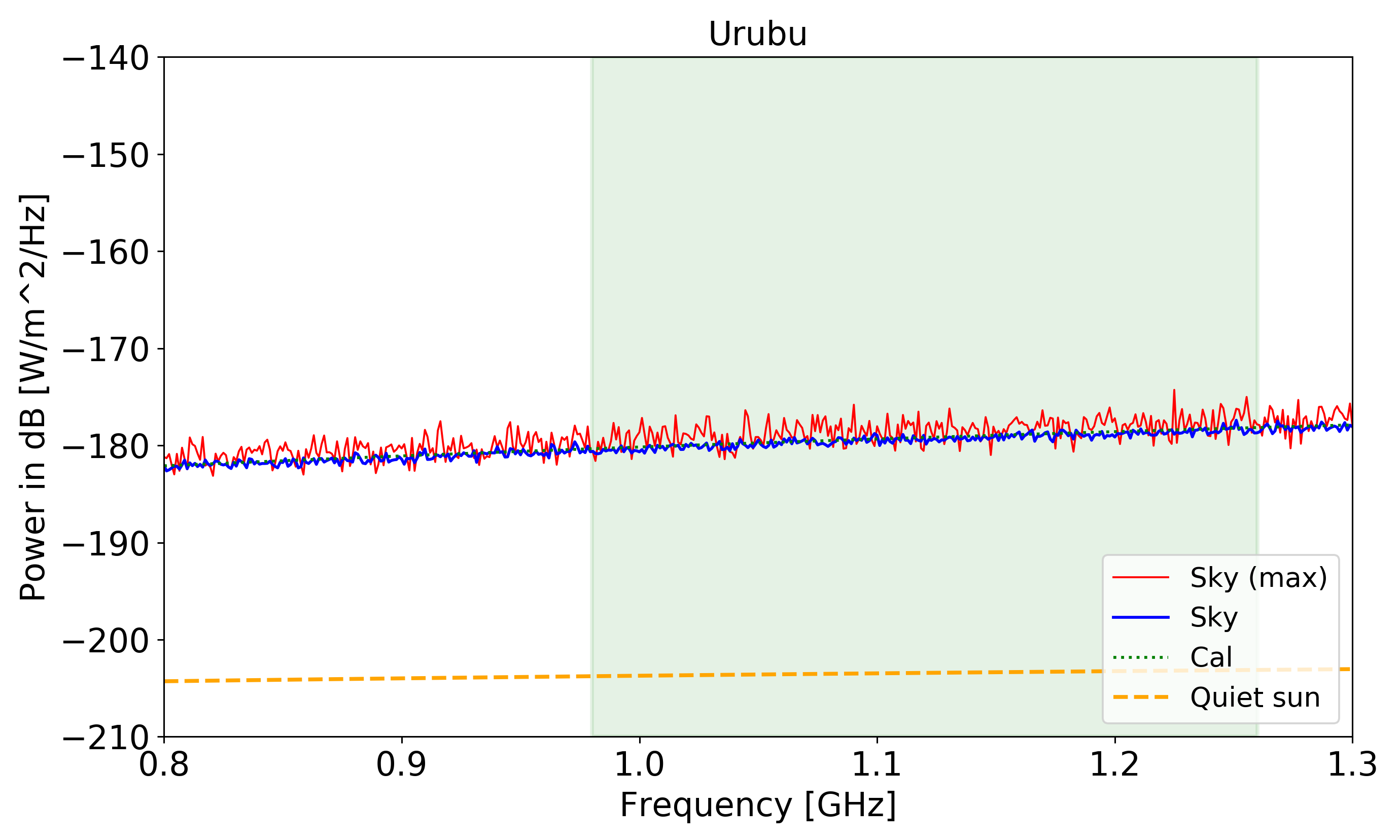}
   \caption{RFI measurements in \paraiba: {\it Parque dos Dinosauros} (top-left), {\it Cruzeiro de Pianc\'o} (top-right), \gato\ (bottom-left) and \urubu\ (bottom-right). Measurements were with a discone antenna with an integration time of 10 minutes. Colours are as per Fig.~\ref{fig:montevideo}. Note the very clean environments in the bottom plots, with a very small mobile phone signal in \gato.}
   \label{fig:RFI_PB}
\end{figure}

We show two example sites from the set of sites that were investigated in \paraiba\ in Figure \ref{fig:RFI_PB}. Parque dos Dinosauros (``B3'' in Table \ref{tab:sites}) was a site at a scientific reserve that was located relatively close to a town, and demonstrates high levels of mobile phone signal and a possible in-band signal. Cruzeiro de Pianc\'o (``B4'' in Table \ref{tab:sites}) was a more isolated site in the west of \paraiba\ that still had RFI at 800--950\,MHz but no in-band signals. Both of these sites (as well as the others that were investigated) did not meet the RFI requirements.

Two very good locations were considered for potential sites in the west of \paraiba, identified as ``B5'' and ``B6'' in Table \ref{tab:sites}. Both are surrounded by small ridges, named \gato\ and \urubu. Access to both sites are through dust roads and there are no mobile phone or other detected signals down to the sensitivity of our equipment. Measurements were carried in 3--4 positions in both locations. The best results for \gato\ and \urubu\ are shown in Fig.~\ref{fig:RFI_PB}. Besides being very clean in terms of RFI, the local topography and local support, distinguished them as ideal sites. Ultimately the site at \urubu\ was selected, mostly due to the local topography and better shielding from the neighborhood.

Windmill parks in \paraiba\ are a potential concern regarding the RFI level at the BINGO site. Two windfarms near the coast (06$^\circ$34\am43\as S 34$^\circ$58\am28\as W), located about 360\,km away from \urubu, and three windfarms in {\it Junco do Serid\'{o}.} area (06$^\circ$52\am26\as S, 36$^\circ$49\am07\as W), at about 120\,km from \urubu, produce energy at a level of tens of MW being a RFI source that can jeopardize the project. However, the earth surface curvature and the natural protection offered by \emph{Serra da Catarina} provide an effective shelter from the current wind farms in \paraiba.

\begin{figure}
 \begin{center}
  \includegraphics[width=0.48\linewidth]{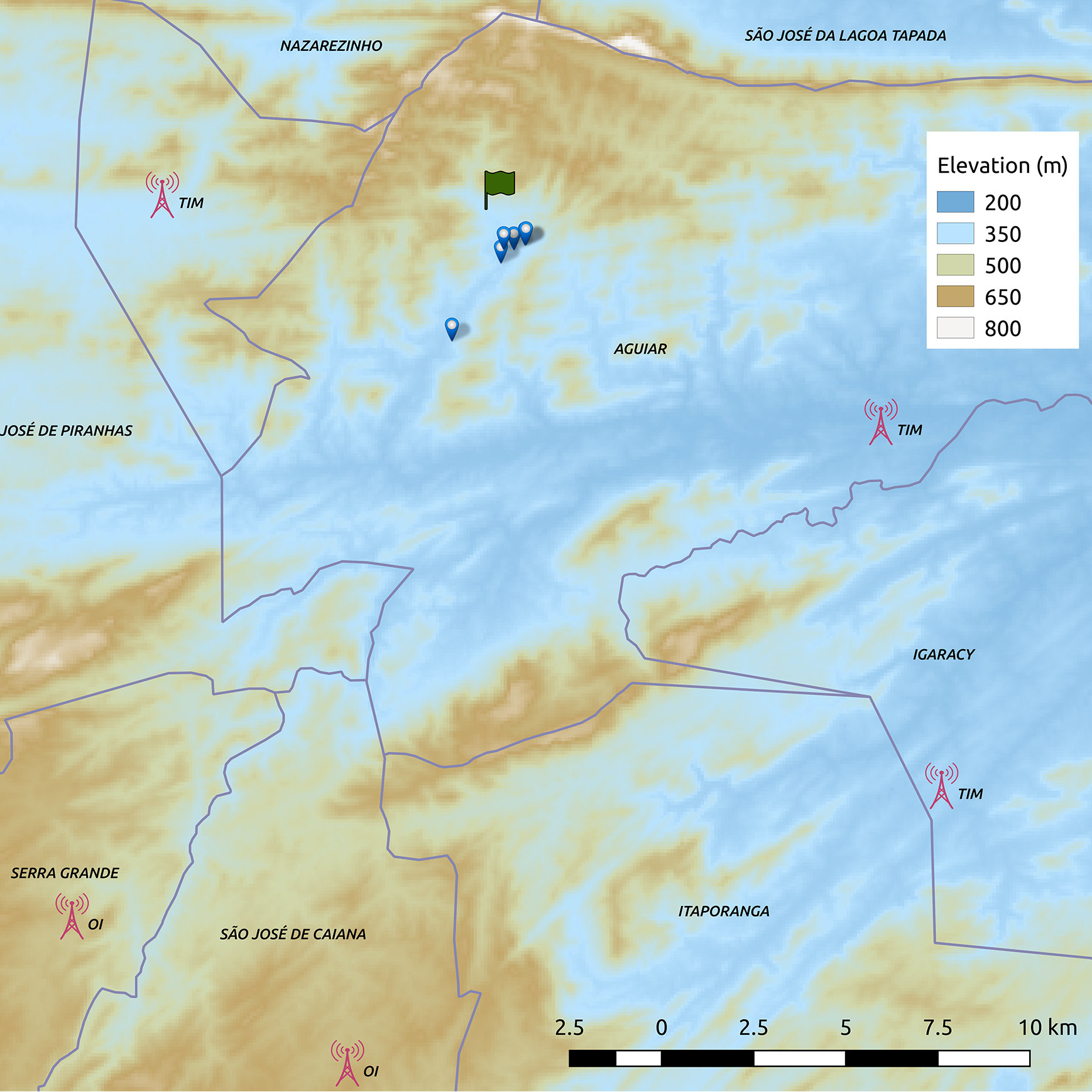}
  \includegraphics[width=0.48\textwidth]{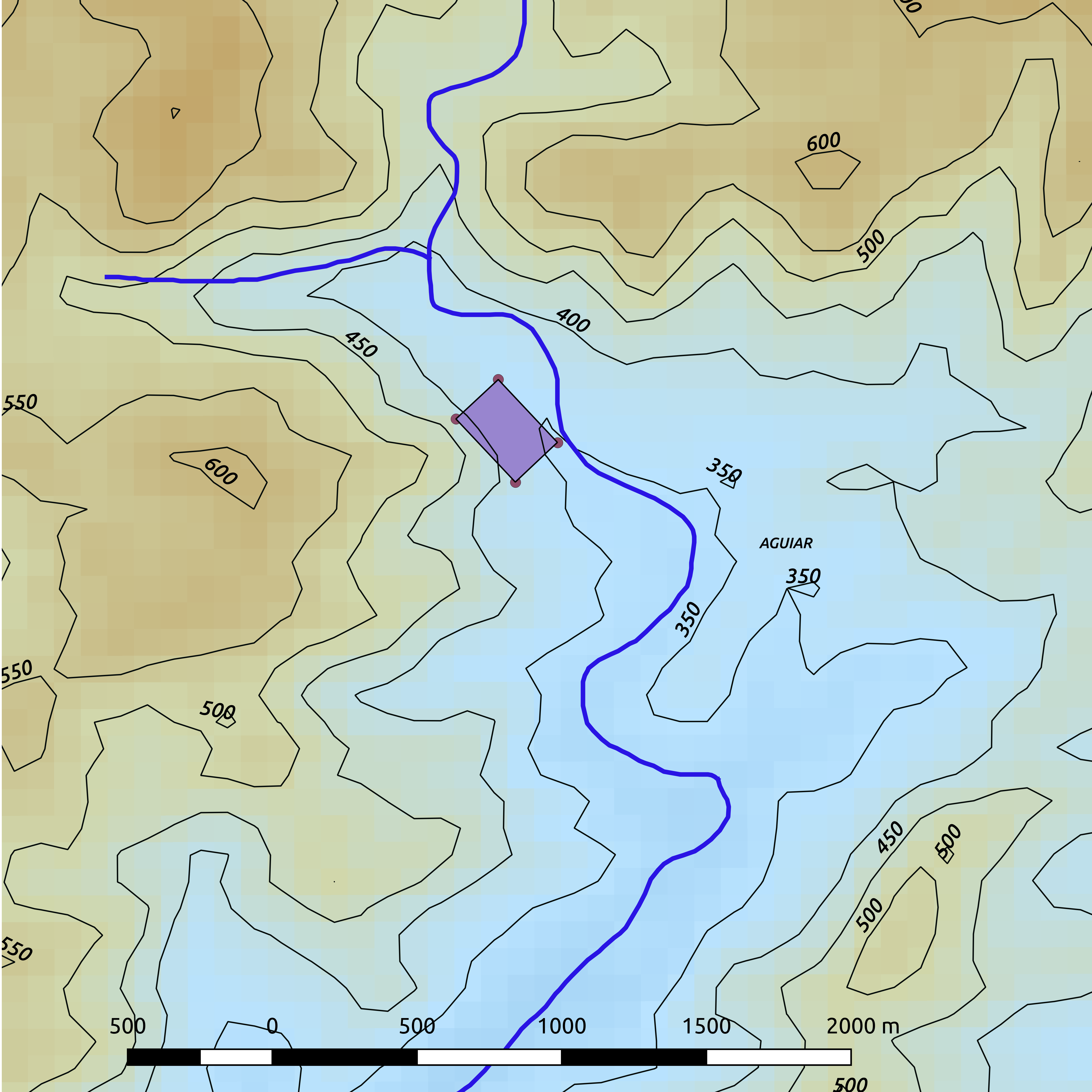}
 \caption{Left Panel: Topographic view of \urubu, including sites from the RFI campaign (markers), nearest mobile antennas with indication of the Telecom and the selected site (flag). Right Panel: Zooming in the BINGO selected site, with contour lines indicating elevation. The thick line indicates an intermittent river. All the elevations were obtained from NASA SRTM data with 90\,m ground resolution.}
  \label{topo.a}
 \end{center}
\end{figure}

The exact telescope location within the site was then subject to further analysis. It is clear from the left panel in Fig.\,\ref{topo.a} that the measured sites are sufficiently shielded for incoming RFI from all nearby antennas, except from the one located in Serra Grande. The measurements did not detect any RFI coming from this antenna, but it was reasonable to search for a site with further protection. The easiest way to use the terrain to assure shielding would be to move northwest, in the direction of the green flag in the left panel in Fig.\,\ref{topo.a}. The goal was to look for additional RFI shielding from the nearby hills, avoiding the possibility of floods and to find a suitably inclined slope to minimize the telescope structure costs. The right panel in Fig.\,\ref{topo.a} shows the selected area for BINGO construction zoomed in. 

The selected terrain has an area of $\approx 200\times300$\,m, sufficient to ensure the construction of the telescope and a security perimeter. There are only two inhabitants within a 1\,km radius, with no electrical power.

\begin{wstable}
\caption{Nearest mobile towers around BINGO site. Positions are depicted in Fig.\,\ref{topo.a}}
\begin{tabular}{lcccc}
\toprule
City & Distance & Type \\
\colrule
Carrapateira &  \hphantom{1}8.6\,km &  2G/3G \\
Aguiar &  12.3\,km  & 2G/3G\\
Igaracy &  20.0\,km & 2G/3G\\
Serra Grande &  22.2\,km & 2G/3G\\
S\~ao Jos\'e de Caiana & 23.5\,km&  2G/3G \\
\botrule
\end{tabular}
\label{tab:towers}
\end{wstable}

\section{ADS-B measurements of aircraft}
\label{sec:aircraft}
\begin{figure}
\centering
\includegraphics[width=0.75\textwidth]{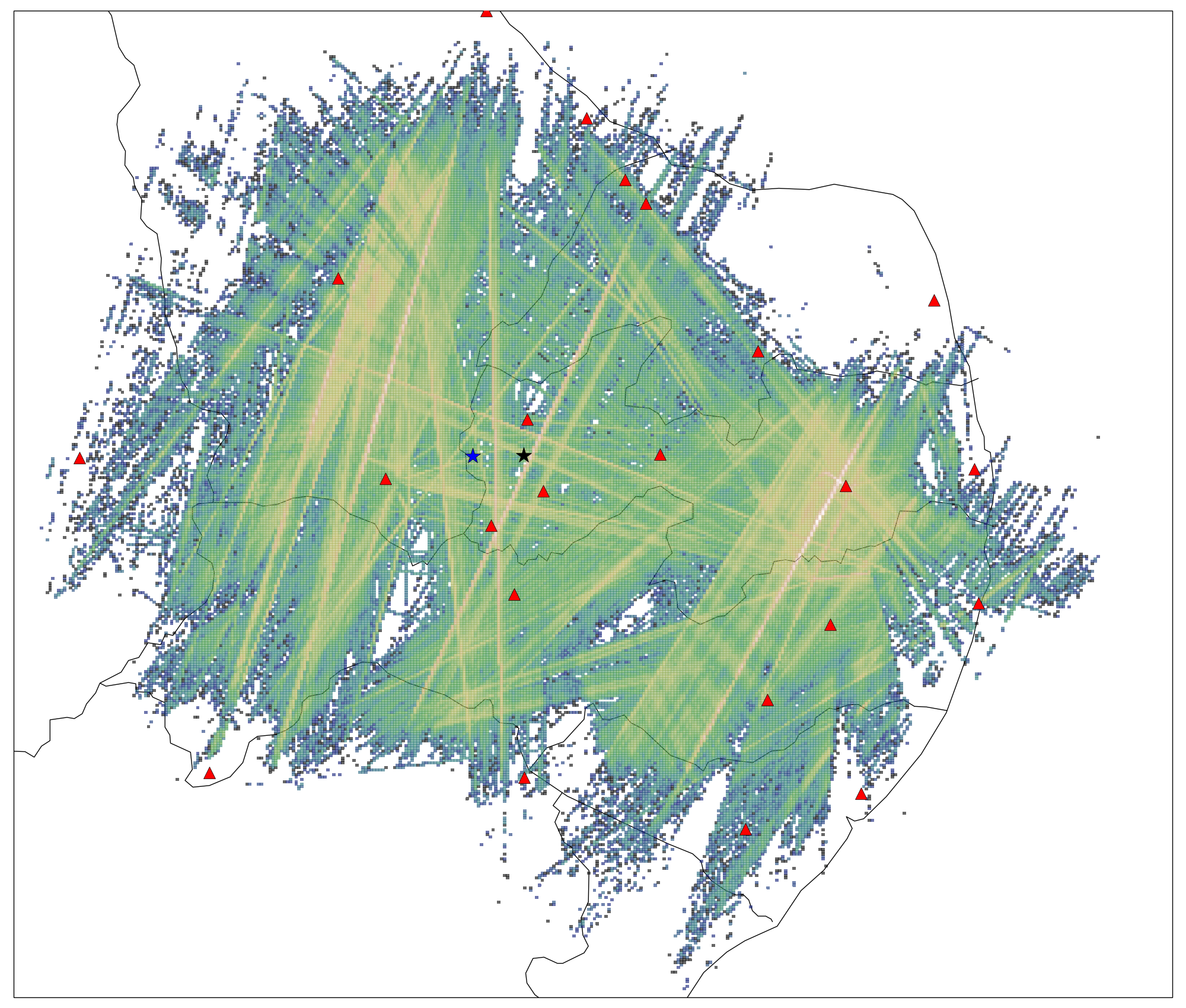}
\caption{A heat map of planes detected using ADS-B measurements in \paraiba. The map covers approximately 600$\times$600\,km. The color scale is logarithmic. \gato\ (left) and \urubu\ (right) are shown as stars. Airports and airfields are shown as triangles.}
\label{fig:adsbmap}
\end{figure}
\begin{figure}
\centering
\includegraphics[width=0.55\textwidth]{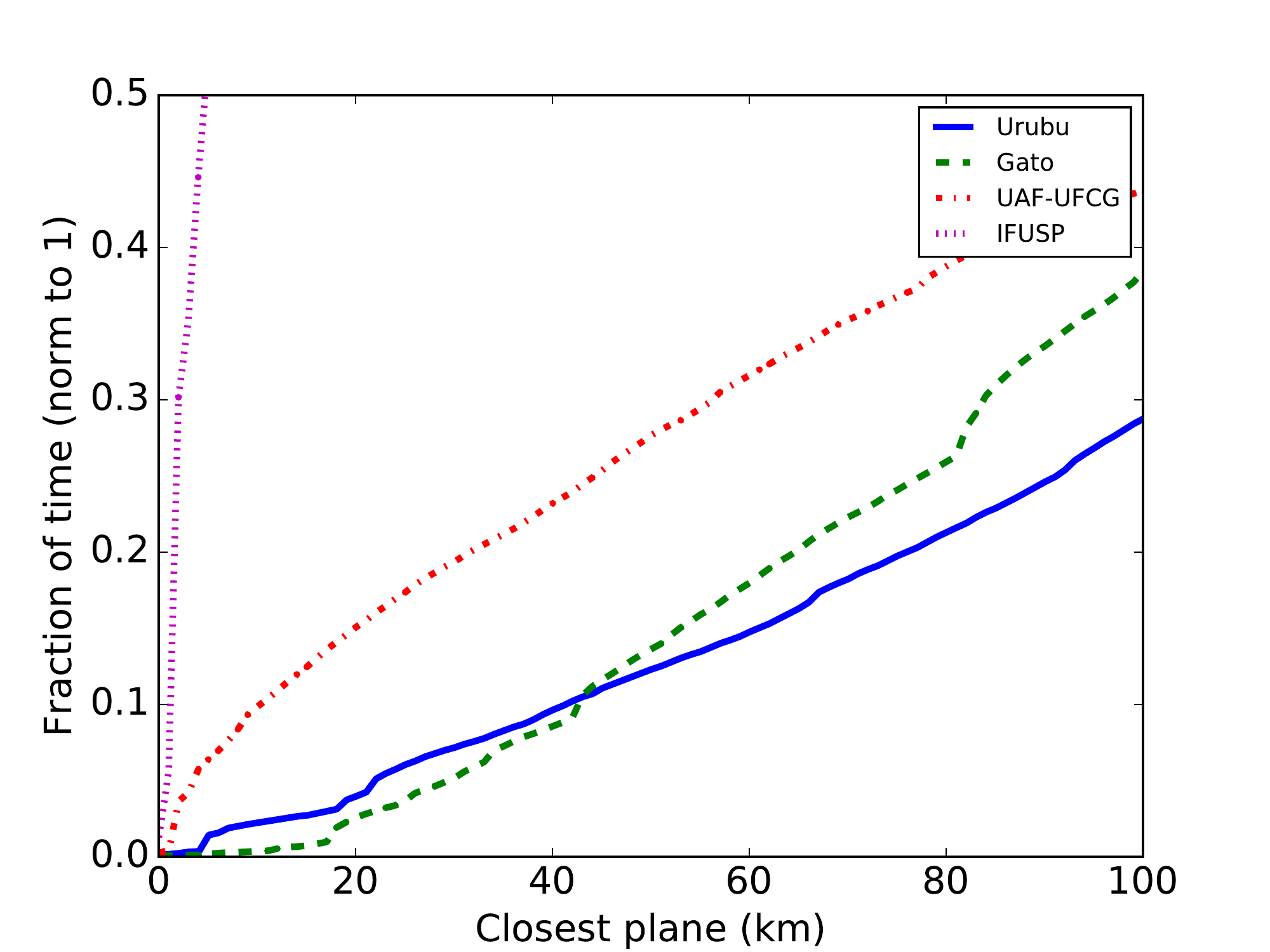}
\caption{Cumulative plots showing the fraction of time that a plane is within a given distance from \urubu\ (blue solid line) and \gato\ (green dashed), with data taken from IFUSP in \saopaulo\ (light blue dots) and UAF-UFCG in Campina Grande (red dash-dot) for comparison.}
\label{fig:cumulative}
\end{figure}

The BINGO frequency band is nominally reserved for communications between airplanes. Automatic dependent surveillance -- broadcast (ADS-B) is an air traffic control signal transmitted by modern airplanes at a frequency of 1090\,MHz. Signals are transmitted very frequently, and include information such as the position, flight number, altitude and heading of the aircraft. Not all aircraft are currently equipped with ADS-B transmitters, but most commercial flights use them. The danger for BINGO, and other radio telescopes, is that the signal is so strong that it could saturate the receiver if the aircraft transmits when in the main beam - and could even damage the receivers.

In order to look at the airplane densities around the potential sites in \paraiba, we installed our own system using \mbox{ADS-B} USB dongles with small 5-inch antennas, attached to raspberry pi 3b's\footnote{\url{https://www.raspberrypi.org/}} running the ``piaware'' software package\footnote{\url{https://flightaware.com/adsb/piaware/}} and a small script to save the ADS-B data to disk. These are the same systems used as a distributed network by various websites that track airplanes around the world, however this network does not provide uniform coverage everywhere, and the fees to access their data are large, which are not an issue with dedicated receivers. We installed two of these in \paraiba, one on a rooftop in Sousa, and another on a rooftop in Campina Grande. Figure \ref{fig:adsbmap} shows the results from 2 months of data overlaid on a map of northern Brazil ($\sim$24 million recorded aircraft positions from late August to late October 2017). This is limited in range to the area shown in dark blue, with common flight paths showing as yellow and white lines.

To work out the fraction of time that we will have an aircraft within a given distance of the site, we use the full dataset to calculate the closest airplane in every three-minute period (sampled every minute) and then plot a normalised histogram of the result, which is shown in Fig.\,\ref{fig:cumulative} for \gato\ and \urubu. From this we can estimate that for 5\% of the time we will have an airplane within $\sim$30\,km of the site, and for 20\% of the time there will be one within $\sim$70\,km. However, this depends on flight routes, which may change in the future.

Since the planes can be tracked, in principle it should be possible to blank the receiver should an aircraft get too close to the telescope, and signals from more distant aircraft can be excised from the signal by removing the affected data channels.

\section{Conclusions for BINGO}\label{sec:conclusions}

\begin{wstable}[t]
\caption{Ranking (best to worst) of the surveyed sites by mobile phone signal level, the presence of in-band signals, and the presence of nearby radar.}
\begin{tabular}{@{}cccccc@{}}
\toprule
rank & \# & Name & Mobile level & In-band signal & Nearby radar \\
\colrule
1 & B6 & \urubu, Brazil & $<-180$\,dB & no & no\\
2 & B5 & \gato, Brazil & $\sim-180$\,dB & no & no\\
3 & U2 & \arerungua, Uruguay & $\sim-170$\,dB & no & no\\
4 & B4 & {\it Cruzeiro de Pianc\'o}, Brazil & $\sim-160$\,dB & no & no\\
5 & U1 & Castrillon, Uruguay & $\sim -150$\,dB & yes & no\\
6 & B3 & {\it Parque dos Dinossauros}, Brazil & $\sim-150$\,dB & possibly & no\\
7 & B1 & S\~ao Martinho, Brazil & $\sim-$150dB & no & yes \\
8 & B2 & Cachoeira Paulista, Brazil & $\sim-$140dB & yes & yes \\
\botrule
\end{tabular}
\label{tab:rank}
\end{wstable}

The RFI measurement campaigns described in this paper have provided a snapshot of the potential RFI on the site. The final ranked list of the surveyed sites shown in this paper is in Table \ref{tab:rank}. \urubu\ is now considered the nominal site for the BINGO telescope, as it had no detectable RFI with a portable receiver during our measurements, given the sensitivity of the equipment.

It is very difficult to compare RFI measurements from different radio observatories, given the variety equipment, methods and units that are used for RFI measurements at the different locations. \urubu\ will be a cleaner site for RFI than most radio observatories, particularly those based in Europe and northern America, although extremely remote observatories such as those for the SKA are likely to be cleaner across a wider frequency range.

The next step will be to set up a permanent RFI monitoring station at \urubu\ to perform longer-term monitoring to identify intermittent sources. Additionally, a radio quiet zone is being defined around the site, with an embargo on any additional mobile phone towers in the area already in place. We have requested a multizone RFI exclusion zone, with no electronic devices permitted within 1\,km of the telescope, restrictions on industrial activities nearby, and restrictions on new transmission towers within $\sim$10\,km of the telescope.

Ultimately, the BINGO telescope itself will be the most sensitive RFI measuring equipment installed on the site. While out-of-band RFI will be filtered out in the receiver, the remaining RFI in the BINGO band will ultimately have to be removed from the data using the data reduction pipeline.

\section*{Acknowledgements}
BINGO is funded by FAPESP (\saopaulo\ State Agency for Reasearch Support) through project \mbox{2014/07885-0}. Partial support of CNPq/Brazil and PEDECIBA/Uruguay is also acknowledged. The RFI campaign in \paraiba\ was partially funded by Universidade Federal de Campina Grande. M.P. acknowledges funding from a FAPESP Young Investigator fellowship, grants 2015/19936-1 and 2016/19425-0. C.A.W. acknowledges support from CNPq through grant 313597/2014-6. S.A. acknowledges support from FAPESP grant 2014/07885-0. C.D. acknowledges support from an ERC Consolidator Grant (no.~307209) and from an STFC Consolidated Grant (ST/P000649/1). K.F. acknowledges FAPESP support through grant 2017/21570-0. T.V. acknowledges CNPq Grant 308876/2014-8. C.A.W. acknowledges the great support from LIT Antenna Testing Facility during the discone characterization. Access to Vale dos Dinosauros (Sousa - PB) was granted by SUDEMA (environmental agency of Paraiba State Government). Data reduction was performed using Python and the numpy, scipy, piaware and cartopy packages. 
\bibliographystyle{ws-jai.bst}
\bibliography{refs}

\end{document}